\documentclass[conference, a4paper]{IEEEtran}

\usepackage{graphicx, epsfig,times,amsmath,amsfonts,amssymb,array,enumerate,algorithm, algorithmic} 

\usepackage[subnum]{cases}
\usepackage[T1]{fontenc}

\newcommand{\bh}{\mathbf{h}}

\newcommand{\bt}{\mathbf{t}}

\newcommand{\bT}{\mathbf{T}}

\newcommand{\bw}{\mathbf{w}}

\newcommand{\bx}{\mathbf{x}}

\newcommand{\bz}{\mathbf{z}}

\newcommand{\bsw}{\boldsymbol{w}}

\newcommand{\cL}{\mathcal{L}}

\newcommand{\bsmu}{\boldsymbol{\mu}}

\newcommand{\bsbeta}{\boldsymbol{\beta}}

\newcommand{\bsSigma}{\boldsymbol{\Sigma}}

\newcommand{\bsPsi}{\boldsymbol{\Psi}}

\newcommand{\Identity}{\textbf{I}}

\newcommand{\E}{\mathbb{E}}

\newcommand{\R}{\mathbb{R}}
	
\newcommand{\N}{\mathcal{N}}

\graphicspath{{Figures/}{Figures/figs-sncf/}}

\IEEEoverridecommandlockouts    

\begin{document}

\title{\ \\ \LARGE\bf Mixture model-based functional discriminant analysis for curve classification \thanks{
Faicel Chamroukhi is with the Information Sciences and Systems Laboratory (LSIS), UMR CNRS  7296 and the University of the South Toulon-Var (USTV). Hervé Glotin is with the LSIS Lab, USTV and Institut Universitaire de France, iuf.amue.fr. Contact: faicel.chamroukhi@univ-tln.fr}}

\author{Faicel Chamroukhi, Herv{\'e} Glotin} 
\maketitle

\begin{abstract}
Statistical approaches for Functional Data Analysis concern the paradigm for which the individuals are functions or curves rather than finite dimensional vectors. In this paper, we particularly focus on the modeling and the classification of functional data which are temporal curves presenting regime changes over time. More specifically, we propose a new mixture model-based discriminant analysis approach for functional data using a specific hidden process regression model. Our approach is particularly adapted to both handle the problem of complex-shaped classes of curves, where each class is composed of several sub-classes, and to deal with the regime changes within each homogeneous sub-class. The model explicitly integrates the heterogeneity of each class of curves via a mixture model formulation,  and the regime changes within each sub-class through a hidden logistic process. The approach allows therefore for fitting flexible curve-models to each class of complex-shaped curves presenting regime changes through an unsupervised learning scheme, to automatically summarize it into a finite number of homogeneous clusters, each of them is decomposed into several regimes. The model parameters are learned by maximizing the observed-data log-likelihood for each class by using a dedicated expectation-maximization (EM) algorithm. Comparisons on simulated data and real data with alternative approaches, including functional linear discriminant analysis and functional mixture discriminant analysis with polynomial regression mixtures and spline regression mixtures, show that the proposed approach provides better results regarding the discrimination results and significantly improves the curves approximation.

\end{abstract}

\section{Introduction}

In many areas of application, such as diagnosis of complex systems \cite{chamroukhi_et_al_neurocomputing2010}\cite{chamroukhi_adac_2011}, electrical engineering \cite{hebrail_etal_Neurocomputing2010}, speech recognition (e.g. the phoneme data studied in \cite{Delaigle2012}), radar waveform \cite{DaboNiang2007}, etc, the data are curves or functions rather than finite dimensional vectors. %
Statistical approaches for Functional Data Analysis (FDA) concern the paradigm of data analysis for which the individuals are entire functions or curves rather than finite dimensional vectors.
 The goals of FDA, as in classical data analysis, include data representation for further analysis,  data visualization, exploratory analysis by performing unsupervised approaches, regression, classification, etc. 
 Additional background on FDA, examples and analysis techniques can be found in \cite{ramsayandsilvermanFDA2005}.  
From a statistical learning prospective, this can be achieved by learning adapted statistical models, in different contexts, e.g., supervised, unsupervised, etc. 
The challenge is therefore to build adapted models to be learned from such data living in a very high or an infinite dimensional space. 
In this paper, we consider the problem of supervised functional data classification (discrimination) where the observations are temporal curves presenting regime changes over time. 
We mainly focus on generative approaches which may help us to understand the process generating the curves. The generative approaches for functional data are essentially based on regression analysis, including polynomial 
regression, splines and B-splines \cite{GaffneyThesis, chamroukhi_PhD_2010, Gui_FMDA, garetjamesANDtrevorhastieJRSS2001}, or also generative polynomial piecewise regression as in \cite{chamroukhi_PhD_2010, chamroukhi_et_al_neurocomputing2010}. Non-parametric statistical approaches have also been proposed for functional data discrimination as in \cite{FerratyV03,Delaigle2012} and clustering as in \cite{Delaigle2012}.
The generative models aim at understanding the process generating such data to handle both the problem of heterogeneity between curves and the process governing the regime changes, in order to fit  flexible models that provide better classification results.
In this paper, we propose a new generative approach for modeling classes of complex-shaped curves where each class is itself composed of unknown homogeneous sub-classes. In addition, the model is particularly dedicated to address the problem when each homogeneous sub-class presents regime changes over time. 
We extend the functional discriminant analysis approach presented in \cite{chamroukhi_et_al_neurocomputing2010}, which relates modeling each class of curves presenting regime changes with a single mean curve, to a mixture formulation which leads to a functional mixture-model based discriminant analysis. More specifically, this approach uses a mixture of regression models with hidden logistic processes (RHLP) \cite{chamroukhi_PhD_2010, chamroukhi_adac_2011} for each class of functional data and derives a functional mixture discriminant analysis framework for functional data classification. The resulting discrimination approach is therefore a model-based functional discriminant analysis in which learning the parameters of each class  of curves is achieved through an unsupervised estimation of a mixture of RHLP (MixRHLP) models.

In the next section we give a brief background on discriminant analysis approaches for functional data  classification  including functional linear and mixture discriminant analysis, 
and then we present the proposed mixture model-based functional mixture discriminant analysis with hidden process regression for curve classification, which we will abbreviate as  FMDA-MixRHLP, and the corresponding parameter estimation procedure using a dedicated expectation-maximization (EM) algorithm.
 
Let us denote by $((\bx_1,y_1),\ldots,(\bx_n,y_n))$ a given labeled training set of curves issued from $G$ classes where $y_i \in  \{1,\ldots,G \}$ is the class label of the $i$th curve $\bx_i$. We assume that $\bx_i$ consists of $m$ observations $(x_{i1},\ldots,x_{im})$, regularly observed at the time points $(t_1,\ldots,t_m)$ with $t_1<\ldots<t_m$. 
 
\section{Background on Functional Discriminant Analysis}
In this section, we give a background on generative discriminant analysis approaches for functional data classification.

Functional discriminant analysis extends discriminant analysis approaches for vectorial data to functional data or curves. From a probabilistic point a view, the conditional density of each class of curves is then assumed to be a (parametric) density defined in the functional space, rather than in a finite dimensional space of the multidimensional data vectors, which is the case for discriminant analysis for vectorial data.
The functional discriminant analysis principle is as follows. Assume we have  a labeled training set of curves and the classes' parameter vectors $(\bsPsi_1,\ldots,\bsPsi_G)$ where $\bsPsi_g$ is the parameter vector of the density of class $g$ $(g=1,\ldots,G)$ (e.g., provided by an estimation procedure from a training set). 
In functional discriminant analysis, a new curve $\bx_i$ is assigned to the class $\hat{y}_i$ using the  maximum a posteriori (MAP) rule, that is: 
\begin{equation}
\hat{y}_i=\arg \max_{1\leq g\leq G} \frac{w_g p(\bx_i|y_i=g,\bt;\bsPsi_g)}{\sum_{g'=1}^{G}w_{g'}p(\bx_i|y_i=g',\bt;\bsPsi_{g'})},
\label{eq: MAP rule for FDA classification}
\end{equation}where $w_g = p(y_i=g)$ is the prior probability of class $g$, which can be computed as the proportion of the class $g$ in the training set, and $p(\bx_i|y_i=g,\bt;\bsPsi_g)$ its  conditional density. 

There are different ways to model this conditional density. 
By analogy to linear or quadratic discriminant analysis for vectorial data, the  class conditional density for each class of curves can be defined as a density of a single model, e.g., a polynomial regression model,  spline, including B-spline \cite{garetjamesANDtrevorhastieJRSS2001}, or a generative piecewise regression model with a hidden logistic process (RHLP)  \cite{chamroukhi_et_al_neurocomputing2010} when the curves further present regime changes over time. These approaches lead to Functional Linear (or quadratic) Discriminant Analysis which we will abbreviate as (FLDA).

The next section briefly recalls the FLDA based on polynomial or spline regression.
\subsection{Functional Linear Discriminant Analysis}
\label{ssec: FLDA state of the art}
Functional Linear (or Quadratic) Discriminant Analysis (FLDA) \cite{garetjamesANDtrevorhastieJRSS2001} arises when we model each  class conditional density of curves with a single model. More specifically, the conditional density $p(\bx_i|y=g,\bt;\bsPsi_g)$ in Equation (\ref{eq: MAP rule for FDA classification}) can for example be the one of a polynomial, spline or B-spline regression model with parameters  $\bsPsi_g$, that is:
\begin{IEEEeqnarray}{lcl}
p(\bx_i|y_i=g,\bt;\bsPsi_g) = \N (\bx_{i};\bT \bsbeta_g,\sigma_g^2\Identity_m), 
\end{IEEEeqnarray}where $\bsbeta_{g}$ is the 
coefficient  vector of the polynomial or spline regression model representing class $g$ and $\sigma_{g}^2$ the associated noise variance, the matrix $\bT$ is the matrix of design which depends on the adopted model (e.g., for polynomial regression, $\bT$ is the $m\times(p+1)$ Vandermonde matrix with rows $(1,t_j,t_j^2,\ldots,t_j^p)$ for $j=1,\ldots,m.$, $p$ being the polynomial degree) and $\N (.;\bsmu,\bsSigma)$ represents the multivariate Gaussian density with mean $\bsmu$ and covariance matrix $\bsSigma$.
 Estimating the model for each class in this case consists therefore in estimating the regression model parameters $\bsPsi_g$ by maximum likelihood which is in this case equivalent to performing least squares estimation. 
A similar FLDA approach that fits a specific generative piecewise regression model governed by a hidden logistic process to 
homogeneous classes of curves presenting regime changes has been presented in \cite{chamroukhi_et_al_neurocomputing2010}.

However, all these approaches, as they involve a single model for each class, are only suitable for homogeneous classes of curves. For complex-shaped classes, when one or more classes are dispersed, the hypothesis of a single model description for the whole class of curves becomes restrictive. This problem can be handled, by analogy to mixture discriminant analysis for vectorial data \cite{hastieANDtibshiraniMDA}, by adopting a mixture model formulation \cite{mclachlanFiniteMixtureModels, titteringtonBookMixtures} in the functional space for each class  of curves. The functional mixture can for example be a polynomial regression mixture 
 or a spline regression mixture 
 \cite{GaffneyThesis, chamroukhi_PhD_2010, Gui_FMDA}. This leads to Functional Mixture Discriminant Analysis (FMDA) \cite{chamroukhi_PhD_2010, Gui_FMDA}.
  
  The next section describes the previous work on FMDA which uses polynomial regression and spline regression mixtures.

\subsection{Functional Mixture Discriminant Analysis with  polynomial regression and spline regression mixtures}
\label{FMDA from the state of the art}
 
A first idea on Functional Mixture Discriminant Analysis (FMDA), motivated by the complexity of the 
time course gene expression functional data for which modeling each class with a single function using FLDA is not adapted, 
was proposed in \cite{Gui_FMDA} and is based on B-spline regression mixtures.  
In the approach of \cite{Gui_FMDA}, each class $g$ of functions is modeled as a mixture of $K_g$ sub-classes, each sub-class  $k$ ($k=1,\ldots,K_g$) is a noisy  B-spline function (can also be a polynomial or  a spline function) with parameters $\bsPsi_{gk}$. 
The model is therefore defined by the following conditional mixture density: 
\begin{IEEEeqnarray}{lcl}
p(\bx_i|y_i = g, \bt; \bsPsi_g) &=& \sum_{k=1}^{K_g} \alpha_{gk} \ p(\bx_i|y_i=g,z_i=k, \bt;\bsPsi_{gk}) \IEEEnonumber \\
&=& \sum_{k=1}^{K_g} \alpha_{gk} \N (\bx_{i};\bT \bsbeta_{gk},\sigma_{gk}^2\Identity_m), 
\label{eq: class mixture density for classic FMDA}
\end{IEEEeqnarray}where the $\alpha_{gk}$'s are the non-negative mixing proportions that sum to 1   such that  $\alpha_{gk} = p(z_i = k|y_i=g)$ ($\alpha_{gk}$ represents the prior probability of the sub-class $k$ of class $g$),
 $z_i$ is a hidden discrete variable in $\{1,\ldots,K_g\}$ representing the labels of the sub-classes for each class.   The parameters of this functional mixture density (Equation (\ref{eq: class mixture density for classic FMDA})) for each class $g$, denoted by 
$$\bsPsi_g = (\alpha_{g1},\ldots,\alpha_{gK_g}, \bsPsi_{g1},\ldots,\bsPsi_{gK_g})$$
can be estimated by maximizing the observed-data log-likelihood by using the expectation-maximization (EM) algorithm \cite{dlr} \cite{mclachlanEM} as in \cite{Gui_FMDA}.

However, using polynomial or spline regression for class representation, as studied in \cite{chamroukhi_PhD_2010, chamroukhi_et_al_neurocomputing2010} is more adapted for curves presenting smooth regime changes and for the splines the knots have to be fixed in advance. 
When the regime changes are abrupt, capturing the regime transition points needs to relax the regularity constraints on splines which leads to piecewise regression for which the knots can be optimized using a dynamic programming procedure. On the other hand, the regression model with a hidden logistic process (RHLP) presented in \cite{chamroukhi_et_al_neurocomputing2010} and used to model each homogeneous set of curves with regime changes, is flexible and explicitly integrates the smooth and/or abrupt regime changes via a logistic process. As pointed in \cite{chamroukhi_et_al_neurocomputing2010}, this approach however has  limitations in the case of complex-shaped classes of curves since each class is only approximated by a single RHLP model.

In this paper, we  extend the discrimination approach proposed in \cite{chamroukhi_et_al_neurocomputing2010} which is based on functional linear discriminant analysis (FLDA) using  a single density model (RHLP) for each class, to a functional mixture discriminant analysis framework (FMDA), where each class conditional density model is assumed to be a mixture of regression models with hidden logistic processes (which we abbreviate as MixRHLP).
Thus, by using this Functional Mixture Discriminant Analysis approach, We may therefore overcome the limitation of FLDA (and FQDA) for modeling complex-shaped classes of curves, via the mixture formulation. Furthermore,  thanks to the flexibility to the RHLP model that approximates each sub-class, as studied in \cite{chamroukhi_et_al_NN2009, chamroukhi_et_al_neurocomputing2010}, we will also be able to automatically and flexibly approximate the underlying hidden regimes.  

The proposed functional mixture discriminant analysis with hidden process regression and the unsupervised learning procedure  for each class through the EM algorithm, are presented in the next section. 

\section{Proposed Functional Mixture Discriminant Analysis with hidden process regression mixture}
\label{FMDA from the state of the art}

Let us assume as previously that each class $g$ $(g=1,\ldots,G)$ has a complex shape so that it is composed of $K_g$ homogeneous sub-classes. Furthermore, now let us suppose that each sub-class $k$ $(k=1,\ldots,K_g)$ of class $g$  is itself governed by $R_{gk}$ unknown regimes. We let therefore $h_{gkj} = r \in \{1,\ldots,R_{gk}\}$ denotes the discrete  variable representing the regime label for sub-class $k$ of class $g$.

\subsection{Modeling the classes of curves with a mixture of regression models with hidden logistic processes} 
 
In the proposed functional mixture discriminant analysis approach, 
we model each class of curves by a specific mixture of regression models with hidden logistic processes (MixRHLP) as in \cite{chamroukhi_PhD_2010,chamroukhi_adac_2011}.  According to the MixRHLP model, each class of curves $g$ is assumed to be composed of $K_g$ homogeneous sub-groups with prior probabilities $\alpha_{g1},\ldots,\alpha_{gK_g}$. Each of the $K_g$ sub-groups is governed by $R_{gk}$ hidden polynomial regimes and is modeled by a regression model with hidden logistic process (RHLP). The RHLP model \cite{chamroukhi_et_al_NN2009, chamroukhi_et_al_neurocomputing2010} assumes that the curves of each sub-class (or cluster) $k$ of class $g$ are   generated by $K_g$  polynomial regression models governed by a hidden logistic process $\bh_{gk}=(h_{gk1},\ldots,h_{gkm})$ that allows for  switching from one regime to another among $R_g$ polynomial regimes over time.   
Thus, the distribution of a curve $\bx_i$ belonging to sub-class $k$ of class $g$ is defined by:
\begin{IEEEeqnarray}{lll}
p(\bx_i|y_i = g, z_i=k,\bt;\bsPsi_{gk})  = && \IEEEnonumber \\
 \prod_{j=1}^m \sum_{r=1}^{R_{gk}}\pi_{gkr}(t_j;\bw_{gk})\mathcal{N}\big(x_{ij};\bsbeta_{gkr}^T \bt_{j},\sigma_{gkr}^{2} \big) 
 \label{eq: RHLP}
\end{IEEEeqnarray}
where $\bsPsi_{gk} = (\bw_{gk},\bsbeta_{gk1},\ldots,\bsbeta_{gkR_{gk}},\sigma^2_{gk1},\ldots,\sigma^2_{gkR_{kg}})$ for $(g=1,\ldots,G; k=1,\ldots,K_g)$
is its parameter vector.  
The quantity $\pi_{gkr}(t_j;\bw_{gk})$ represents the  probability of regime $r$ within sub-class $k$ of class $g$ and is modeled by a logistic distribution, that is:  
\begin{IEEEeqnarray}{lcl}
 \pi_{gkr}(t_j;\bw_{gk}) &=& p(h_{gkj}=r|t_j;\bw_{gk}) \IEEEnonumber \\
 &=& \frac{\exp{(w_{gkr0} + w_{gk1}t_j)}}{\sum_{\ell=1}^{R_{gk}}\exp{(w_{g\ell r 0} + w_{g \ell r 1} t_j)}},
\label{eq: logistic prob for regime g k r}
\end{IEEEeqnarray}where $\bw_{gk} = (\bsw_{gk1},\ldots,\bsw_{gkR_{gk}})$ is its parameter vector, $\bsw_{gkr}=(w_{gkr0},w_{gkr1})^T$ being the $2$-dimensional coefficient vector for the $r$th logistic component. The hidden process $\bh_{gk}$ governing each sub-class is therefore assumed to be logistic. The relevance of the logistic process in terms of flexibility of transitions has been well detailed in \cite{chamroukhi_et_al_NN2009, chamroukhi_et_al_neurocomputing2010}.

Thus, the resulting conditional distribution of a curve $\bx_i$ issued from class $g$ is given by the following conditional mixture density: 
{\small \begin{IEEEeqnarray}{lcl}
p(\bx_i|y_i\! =\! g, \bt;\bsPsi_g) \!\! = \!\! \sum_{k=1}^{K_g} p(z_i \! = \! k|y_i \! =\! g) p(\bx_i|y_i\! =\! g, z_i\! =\! k,\bt;\bsPsi_{gk}) & &\IEEEnonumber \\
= \sum_{k=1}^{K_g} \alpha_{gk} \prod_{j=1}^m \sum_{r=1}^{R_{gk}}\pi_{gkr}(t_j;\bw_{gk})\mathcal{N}\big(x_{ij};\bsbeta_{gkr}^T \bt_{j},\sigma_{gkr}^{2} \big) &&
\label{eq: MixRHLP}
\end{IEEEeqnarray}}where $\bsPsi_g =(\alpha_{g1},\ldots,\alpha_{gK_g},\bsPsi_{g1},\ldots,\bsPsi_{gK_g})$ is the parameter vector for class $g$,  
 $\bsPsi_{gk}$, 
 being the parameters of each of its RHLP component density $\prod_{j=1}^m \sum_{r=1}^{R_{gk}}\pi_{gkr}(t_j;\bw_{gk})\mathcal{N}\big(x_{ij};\bsbeta_{gkr}^T \bt_{j},\sigma_{gkr}^{2} \big)$ as given by Equation (\ref{eq: RHLP}). Notice that the key difference between the proposed FMDA with hidden process regression and the FMDA proposed in \cite{Gui_FMDA} is that the proposed approach uses a generative hidden process regression model (RHLP) for each sub-class rather than a spline; the RHLP is itself based on a mixture formulation.  Thus, the proposed approach is more adapted for capturing the regime changes within curves.  

Now, once we have defined the model for each class of curves $g$, we have to estimate its parameters $\bsPsi_g$. The next section presents the unsupervised learning of the model parameters $\bsPsi_g$ for each class of curves by maximizing the observed-data log-likelihood through the EM algorithm.

\subsection{Maximum likelihood  estimation via the EM algorithm}
\label{sec: parameter estimation by EM mixture functional rhlp} 

Given an independent training set of labeled curves, the parameter vector $\bsPsi_g$ of the mixture density of class $g$ given by Equation (\ref{eq: MixRHLP}) is estimated by maximizing the following observed-data log-likelihood:
{\small \begin{IEEEeqnarray}{lcl} 
\!\! \cL(\bsPsi_g) & = & \log \!\! \prod_{i|y_i =g}\!\! p(\bx_i|y_i \! =\! g, \bt;\bsPsi_g)\IEEEnonumber\\
&\!\! =\!\! & \!\!\sum_{i|y_i=g}\!\!\! \log \! \sum_{k=1}^{K_g} \alpha_{gk} \! \prod_{j=1}^m \! \sum_{r=1}^{R_{gk}} \!\! \pi_{gkr}(t_j;\bw_{gk})\mathcal{N}\big(x_{ij};\bsbeta_{gkr}^T \bt_{j},\sigma_{gkr}^{2} \big).\IEEEnonumber
\label{eq: loglik MixFRHLP for class g}
\end{IEEEeqnarray}}The maximization of this log-likelihood cannot be performed in a closed form. We maximize it 
iteratively by using a dedicated EM algorithm. 
The EM scheme requires the definition of the complete-data log-likelihood. The complete-data log-likelihood for the proposed MixRHLP model for each class, given the observed data which we denote by $\mathcal{D} = (\{\bx_i|y_i=g\},\bt)$, the hidden cluster labels $\bz=(z_1,\ldots,z_n)$, and the hidden processes $\bh_{gk}=(h_{1gk},\ldots,h_{mgk})$, governing  each of the $K_g$ clusters, is given by:
\vspace*{-.08cm}
{\small \begin{IEEEeqnarray}{lll}
 \cL_c(\bsPsi_g) & \! = \! &\sum_{i|y_i\! =\! g}\! \sum_{k=1}^{K_g} \! z_{ik}\! \Big[ \log \alpha_{gk} \! + \!\!\! \sum_{j=1}^{m}\sum_{r=1}^{R_{gk}} \! h_{jgkr} \! \log \pi_{gkr}(t_j;\bw_{gk}) \IEEEnonumber \\
&  & +  \sum_{j=1}^{m} \sum_{r=1}^{R_{gk}} h_{jgkr} \log \mathcal{N}\left(y_{ij};{\bsbeta}^{T}_{gkr}\bt_{j},\sigma^2_{gkr}\right)\Big]. 
\label{eq: complete log-lik for the MixRHLP}
\end{IEEEeqnarray}}where  $z_{ik}$ and $h_{jgkr}$ are indicator binary-valued variables such that $z_{ik}=1$ if $z_i=k$ (i.e., if the $i$th curve $\bx_i$ is generated by the cluster (sub-class) $k$)  and $z_{ik}=0$ otherwise; and $h_{jgkr}=1$ if $h_{gk}=r$ (i.e.,  the $i$th curve belongs to the sub-class $k$ and its $j$th point $x_{ij}$ belongs to the $r$th regime), and $h_{jgkr}=0$ otherwise.  

The next paragraph shows how the observed-data log-likelihood $\cL(\bsPsi_g)$ is maximized by the EM algorithm.

\subsection{The dedicated EM algorithm for the unsupervised learning of the parameters of the MixRHLP model for each class}
\label{ssec. EM algorithm for mixture functional rhlp}
For each class $g$, the EM algorithm starts with an initial parameter $\bsPsi_g^{(0)}$ and alternates between the two following steps until convergence:

\subsubsection{E-step}
\label{par: E-step mixture of rhlp and EM}
This step computes the expected complete-data log-likelihood, given the observations $\mathcal{D}$, and the current parameter estimation  $\bsPsi_g^{(q)}$, $q$ being the current iteration number: 
{\small \begin{IEEEeqnarray}{lcl} 
\!\! & &\!\!\!\! Q(\bsPsi_g,\bsPsi_g^{(q)}) \! = \! \E\left[\cL_c(\bsPsi_g;\mathcal{D},\bz,\{\bh_{gk}\})|\mathcal{D};\bsPsi_g^{(q)}\right]\IEEEnonumber \\  
\!\! & = &\!
\sum_{i|y_i=g}\!\sum_{k=1}^{K} \!\!  \gamma_{igk}^{(q)} \log \alpha_{gk} \! + \!\!\!\! \sum_{i|y_i=g} \! \sum_{k=1}^{K_g}\! \sum_{j=1}^{m}\! \sum_{r=1}^{R_{gk}}\!\! \gamma_{igk}^{(q)}\tau^{(q)}_{ijgkr}\log \pi_{gkr}(t_j;\bw_{gk}) \IEEEnonumber \\ 
& &  + \!\! \sum_{i|y_i=g} \sum_{k=1}^{K_g} \sum_{j=1}^{m}\sum_{r=1}^{R_{gk}} \gamma_{igk}^{(q)} \tau^{(q)}_{ijgkr}\log \mathcal{N} \left(x_{ij};{\bsbeta}^{T}_{gkr}\bt_{j},\sigma^2_{gkr} \right).\label{eq: Q-function for the MixRHLP for class g}
\end{IEEEeqnarray}}As shown in the expression of $Q(\bsPsi_g,\bsPsi_g^{(q)})$, this step simply requires the calculation of the posterior sub-class probabilities (i.e., the probability that the observed curve $\bx_{i}$ originates from sub-class (cluster) $k$ for class $g$)
{\small \begin{IEEEeqnarray}{lcl}
\gamma_{igk}^{(q)} & = & p(z_{i}=k|\bx_{i},y_i=g,\bt;\bsPsi_{gk}^{(q)}) \IEEEnonumber \\
&=& \frac{\alpha_{gk}^{(q)} p(\bx_i | y_i=g,z_i=k,\bt;\bsPsi^{(q)}_{gk})}{ \sum_{l =1}^{K_g} \alpha_{gl}^{(q)}p(\bx_i |y_i=g, z_i=l,\bt;\bsPsi^{(q)}_{gl})} \IEEEnonumber \\ 	
&=& \frac{\alpha_{gk}^{(q)}\prod_{j=1}^m\sum_{r=1}^{R_{gk}}\pi_{gkr}(t_j;\bw_{gk}^{(q)})\mathcal{N}\big(x_{ij};\bsbeta^{T(q)}_{gkr}\bt_{j},\sigma^{2(q)}_{gkr}\big)}
{ \sum_{l=1}^{K_g} \alpha_{gl}^{(q)}\prod_{j=1}^m\sum_{r=1}^{R_{gl}} \pi_{glr}(t_j;\bw_{gl}^{(q)}) \mathcal{N}(x_{ij};\bsbeta^{(q)T}_{glr}\bt_{j},\sigma^{2(q)}_{glr})} \IEEEnonumber \\
\label{eq: curves post prob gamma_ijgk}
\end{IEEEeqnarray}}and the posterior regime probabilities for each sub-class (i.e., the probability that the observed data point $x_{ij}$ at time $t_j$ originates from the $r$th regime of sub-class $k$ for class $g$), given by:
\begin{IEEEeqnarray}{lcl}
\tau^{(q)}_{ijgkr} &=& p(h_{jgk}=r|x_{ij},y_i=g, z_i=k,  t_j;\bsPsi^{(q)})\IEEEnonumber \\ 
&=&\frac{\pi_{gkr}(t_j;\bw_{gk}^{(q)})\mathcal{N}(x_{ij};\bsbeta^{T(q)}_{gkr}\bt_{j},\sigma^{2(q)}_{gkr})}
{\sum_{\ell=1}^{R_{gk}}\pi_{gk \ell}(t_j;\bw_{gk}^{(q)})\mathcal{N}(x_{ij};\bsbeta^{T(q)}_{gk \ell}\bt_{j},\sigma^{2(q)}_{gk \ell})}\cdot
\label{eq: post prob tau^r_ijk of the segment k of the cluster r for the MixRHLP}
\end{IEEEeqnarray} 

\subsubsection{M-step}
\label{par: M-step mixture of rhlp and EM}
This step updates the value of the parameter $\bsPsi_g$ by maximizing the function $Q(\bsPsi_g,\bsPsi_g^{(q)})$ given by Equation (\ref{eq: Q-function for the MixRHLP for class g}) with respect to $\bsPsi_g$, that is: 
$$\bsPsi_g^{(q+1)} = \arg \max_{\bsPsi_g} Q(\bsPsi_g,\bsPsi_g^{(q)}).$$  
It can be shown that this maximization can be performed by separate  maximizations w.r.t the mixing proportions $(\alpha_{g1},\ldots,\alpha_{gK_g})$ subject to the constraint $\sum_{k=1}^{K_g} \alpha_{gk} = 1$, 
and w.r.t the regression parameters $\{\bsbeta_{gkr},\sigma^2_{gkr}\}$ and the hidden logistic process parameters $ \{\bw_{gk}\}$. 

The mixing proportions updates are given, as in the case of standard mixtures, by 
 \begin{IEEEeqnarray}{lcl}
\alpha_{gk}^{(q+1)} &=& \frac{1}{n_g}\sum_{i|y_i=g} \gamma_{igk}^{(q)},\quad (k=1,\ldots,K_g), 
\label{eq: EM estimate of the cluster prior prob alpha_r for the MixRHLP}
\end{IEEEeqnarray} $n_g$ being the cardinal number of class $g$. 
The maximization w.r.t the regression parameters 
 consists in performing separate analytic solutions of weighted least-squares problems where the weights are the product of the posterior probability $\gamma^{(q)}_{igk}$ of sub-class $k$ and the posterior probability $\tau^{(q)}_{ijgkr}$ of regime $r$ of sub-class $k$. Thus, the regression coefficients updates are given by:
{\small \begin{IEEEeqnarray}{lcl}
\!\!\!\!\!\!\!\!\!\! \bsbeta^{(q+1)}_{gkr} & \!  =  \! &\Big[\! \sum_{i|y_i=g}\! \sum_{j=1}^{m} \gamma_{igk}^{(q)} \tau^{(q)}_{ijgkr}\bt_j \bt_j^T\Big]^{-1} \!\!\! \sum_{i|y_i\! =\! g}\! \sum_{j=1}^{m} \gamma_{igk}^{(q)} \tau^{(q)}_{ijgkr}x_{ij} \bt_j   
\label{eq: EM estimate of reg coeff beta_rk of polynom k of the sub-class r for the MixRHLP}
\end{IEEEeqnarray}
and the updates for the variances are given by:
\begin{IEEEeqnarray}{lcl}
\!\!\!\!\!\! \sigma_{gkr}^{2(q+1)} & = & \frac{\sum_{i|y_i=g}\sum_{j=1}^{m} \gamma_{igkr}^{(q)}\tau^{(q)}_{ijgkr} (x_{ij}-{\bsbeta}^{T(q+1)}_{gkr} \bt_j)^2}{\sum_{i|y_i=g}\sum_{j=1}^m \gamma_{igkr}^{(q)}\tau_{ijgkr}^{(q)} } \cdot  
\label{eq: EM estimate of variance sigma^2_rk for polynom k of the sub-class r for the MixRHLP}
\end{IEEEeqnarray}}Finally, the maximization w.r.t the logistic processes parameters $\{\bw_{gk}\}$ consists in solving multinomial logistic regression problems weighted by $\gamma_{igk}^{(q)}\tau^{(q)}_{ijgkr}$ which we solve with a multi-class IRLS algorithm (e.g., see  \cite{chamroukhi_PhD_2010}). A single update of the IRLS algorithm at iteration $l$ is given by:
{\small \begin{equation}
\!\!{\bw}_{gk}^{(l+1)}\! =\! {\bw}_{gk}^{(l)}-\Big[\frac{\partial^2 Q_{\bw_{gk}})}{\partial \bw_{gk} \partial {\bw_{gk}}^T}\Big]^{-1}_{\bw_{gk}=\bw_{gk}^{(l)}} \frac{\partial Q_{\bw_{gk}}}{\partial \bw_{gk}}\Big|_{\bw_{gk}=\bw_{gk}^{(l)}}.
\label{eq: IRLS functional rhlp}
\end{equation}}where $Q_{\bw_{gk}}$ denotes the terms in the $Q$-function (\ref{eq: Q-function for the MixRHLP for class g}) that depend on $\bw_{gk}$.

The pseudo code \ref{algo: proposed EM algorithm for mixture of functional rhlps} summarizes the EM algorithm for the proposed MixRHLP model.
\begin{algorithm}
\caption{\label{algo: proposed EM algorithm for mixture of functional rhlps} Pseudo code of the proposed algorithm for the MixRHLP model for a set of curves.} 
{\bf Inputs:} Labeled training set of $n$ curves $((\bx_1,y_1),\ldots,(\bx_n,y_n))$ sampled at the time points $\bt=(t_1,\ldots,t_m)$, the number of sub-classes (clusters) $K_g$ ($g=1,\ldots,G$), the number of polynomial regimes $R_{gk}$ and the polynomial degree $p$.
\begin{algorithmic}[1]
\STATE \textbf{Initialize:} $\bsPsi_g^{(0)}= (\alpha^{(0)}_{g1},\ldots,\alpha^{(0)}_{gK_g},\bsPsi_{g1}^{(0)},\ldots,\bsPsi_{gK_g}^{(0)})$ 
\STATE fix a threshold $\epsilon>0$ (e.g., $\epsilon=10^{-6}$), 
\STATE set $q \leftarrow 0$ (EM iteration) 
\WHILE {increment in log-likelihood $> \epsilon$}
\STATE \begin{verbatim}
// E-Step
\end{verbatim} 
	   \FOR{$k=1,\ldots,K_g$}		
	   	\STATE compute $\gamma_{igk}^{(q)}$ for $i=1,\ldots,n$ using Equation (\ref{eq: curves post prob gamma_ijgk})

			\FOR{$r=1,\ldots,R_{gk}$}
				\STATE compute $\tau_{ijgkr}^{(q)}$ for $i=1,\ldots,n$ and $j=1,\ldots,m$ using Equation (\ref{eq: post prob tau^r_ijk of the segment k of the cluster r for the MixRHLP})
			\ENDFOR
	 \ENDFOR
\STATE \begin{verbatim}
// M-Step
\end{verbatim} 
	\FOR{$k=1,\ldots,K_g$}	
		   	\STATE compute the update $\alpha_{gk}^{(q+1)}$ using Equation (\ref{eq: EM estimate of the cluster prior prob alpha_r for the MixRHLP})	
		\FOR{$r=1,\ldots,R_{gk}$}
			\STATE compute the update $\bsbeta_{gkr}^{(q+1)}$ using Equation (\ref{eq: EM estimate of reg coeff beta_rk of polynom k of the sub-class r for the MixRHLP})
			\STATE compute the update $\sigma_{gkr}^{2(q+1)}$ using Equation (\ref{eq: EM estimate of variance sigma^2_rk for polynom k of the sub-class r for the MixRHLP})
		\ENDFOR
		\STATE  \begin{verbatim}
		//IRLS updating loop (Eq. (14))
		\end{verbatim}
	\STATE$\bw_{gk}^{(q+1)} \leftarrow \bw_{gk}^{(l)}$
	\STATE $q \leftarrow q+1$
	  \ENDFOR
	\ENDWHILE
\STATE $\hat{\bsPsi}= (\alpha^{(q)}_{g1},\ldots,\alpha^{(q)}_{gK_g}, \bsPsi^{(q)}_{g1},\ldots \bsPsi^{(q)}_{gK_g})$\end{algorithmic}
{\bf Output:} $\hat{\bsPsi}$ the maximum likelihood estimate of $\bsPsi$
\end{algorithm}

\subsection{Curve classification and approximation with the FMDA-MixRHLP approach} 
\label{ssec: FMDA-MixRHLP classification and approximation}
Once we have an estimate $\hat{\bsPsi}_{g}$ of the parameters of the functional mixture density MixRHLP (provided by the EM algorithm) for each class, a new curve $\bx_i$ is then assigned to the class maximizing the posterior probability (MAP principle) using Equation (\ref{eq: MAP rule for FDA classification}). This therefore leads us to the functional mixture discriminant analysis classification rule (FMDA-MixRHLP) which is particularly adapted to deal with the problem of classes composed of several sub-classes and to further handle the problem of regime changes within each sub-class. 
Regarding to curves approximation, each sub-class $k$ of class $g$ is summarized by approximating it by a single ``mean" curve, which we denote by $\hat{\bx}_{gk}$. Each point $\hat{x}_{gkj}   \ (j=1\ldots,m)$ of this mean curve is defined by the conditional expectation \linebreak $\hat{x}_{gkj}=\E[x_{ij}|y_i=g,z_i=k,t_j;\hat{\bsPsi}_{gk}]$ 
given by:  
\begin{IEEEeqnarray}{lcl} 
\hat{x}_{gkj} &=& \int_{\R}x_{ij} p(x_{ij}|y_i=g,z_i=k,t_j;\hat{\bsPsi}_{gk})dx_{ij} \IEEEnonumber \\
&=&\int_{\R}x_{ij} \sum_{k=1}^K \pi_{gkr}(t_j;\hat{\bw}_{gk}) \mathcal{N}\big(x_{ij};\hat{\bsbeta}^T_{gkr}\bt_{j},\hat{\sigma}^2_{gkr}\big) dx_{ij} \IEEEnonumber \\
&=& \sum_{r=1}^{R_{gk}} \pi_{gkr}(t_j;\hat{\bw}_{gk})\hat{\bsbeta}^T_{gkr} \bt_{j} 
\label{eq: RHLP mean curve}
\end{IEEEeqnarray}which is a sum of polynomials weighted by the logistic probabilities $\pi_{gkr}$ that model the regime variability over time.

\subsection{Model selection}

The number of sub-classes (clusters) $K_g$ for each class $g$ $(g=1,\ldots,G)$ and the number regimes $R_{gk}$ for each sub-class can be computed by maximizing some information criteria e.g., the Bayesian Information Criterion (BIC) \cite{BIC}:
\begin{equation}
\mbox{BIC}(K,R,p)=\cL(\hat{\bsPsi_g})-\frac{\nu_{\bsPsi_g}}{2} \log(n),
\label{eq: BIC for MixFRHLP}
\end{equation}
where $\hat{\bsPsi_g}$  is the maximum likelihood estimate of the parameter vector $\bsPsi_g$ provided by the EM algorithm, $\nu_{\bsPsi_g} = K_g-1 + \sum_{k=1}^{K_g} \nu_{\bsPsi_{gk}}$ is the number of free parameters of the MixRHLP model, $K_g-1$ being the number of mixing proportions and $\nu_{\bsPsi_{gk}} = (p+4)R_{gk}-2$  represents the number of free parameters of each RHLP model associated with sub-class $k$, and $n$ is the sample size. 

\section{Experimental study}  
\label{sec: Experiments} 

This section is dedicated to the evaluation of the proposed approach on simulated data, the waveform benchmark curves of Breiman \cite{breiman} and real data from a railway diagnosis application \cite{chamroukhi_et_al_NN2009,chamroukhi_et_al_neurocomputing2010,chamroukhi_adac_2011}.

 We perform comparisons with alternative functional discriminant analysis approaches  using a polynomial regression (PR) or a spline regression (SR) model \cite{garetjamesANDtrevorhastieJRSS2001}, and the one that uses a single RHLP model as in \cite{chamroukhi_et_al_neurocomputing2010}. These alternatives will be abbreviated FLDA-PR, FLDA-SR and FLDA-RHLP, respectively.  We also consider alternative functional mixture discriminant analysis approaches that use polynomial regression mixtures (PRM), and spline regression mixtures (SRM) as in \cite{Gui_FMDA} which will be abbreviated as FMDA-PRM and FMDA-SRM respectively.

 We use two criteria of evaluation. The first one is the misclassification error rate  computed by a $5$-fold cross-validation procedure and concerns the performance of the approaches in terms of curve classification. The second one is the mean square error between the observed curves  and the estimated mean curves, which is equivalent to the intra-class inertia, and the regards the the performance of the approaches regarding the curves modeling and approximation. For FLDA, as each class $g$ is approximated by a single mean curve $\hat{\bx}_{g}$, this error criterion is therefore given by $\sum_{g}\sum_{i|y_i=g} \parallel \bx_{i} - \hat{\bx}_{g}\parallel^2$, while for FMDA, each class $g$ is summarised by several ($K_g$) mean curves $\{\hat{\bx}_{gk} \}$, each of them summarises a sub-class $k$, and the intra-class inertia in this case is therefore given by $\sum_{g}\sum_{i|y_i=g}\sum_{k=1}^{K_g}\parallel \bx_{i} - \hat{\bx}_{gk}\parallel^2$. Notice that each point of the estimated mean curve for each sub-class is given by a polynomial function or a spline function for the case of polynomial regression mixture or spline regression mixture respectively, or by Equation (\ref{eq: RHLP mean curve}) for the case of the MixRHLP model.  

\subsubsection{Experiments on simulated curves} 
\begin{figure}[!h]
 \centering
 \includegraphics[width=5.6cm, height=3.5cm]{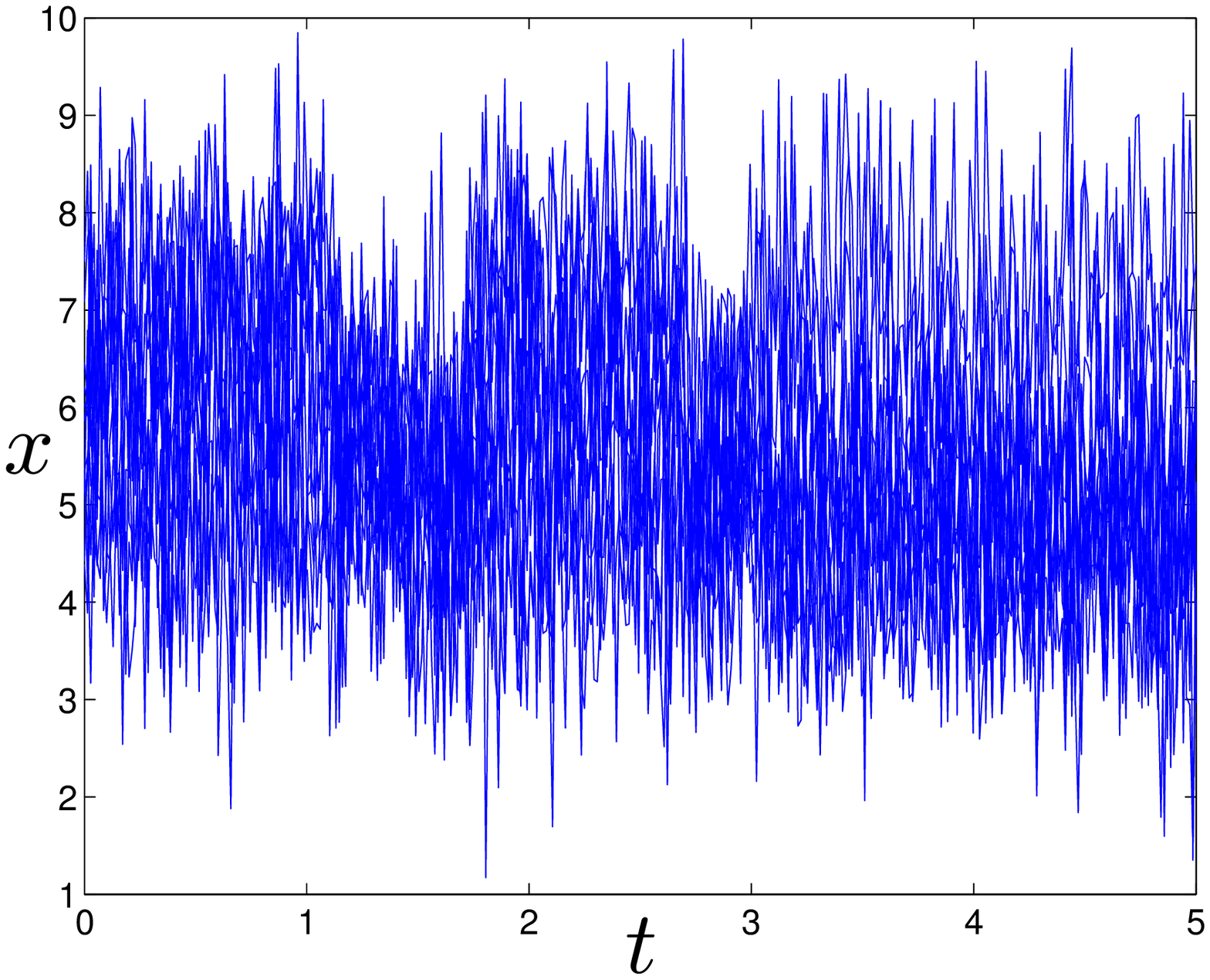}  
 \caption{\label{fig: simulated complex shaped class}
Simulated curves from a complex-shaped class composed of three sub-classes, each of them is composed of three piecewise constant regimes.}
\end{figure}

In this section, we consider simulated curves issued from two classes of piecewise noisy functions. The first class has a complex shape as it is composed of three sub-classes (see Figure \ref{fig: simulated complex shaped class}), while the second one is a homogeneous class. Each  curve consists of three piecewise regimes and is composed of $200$ points.

\begin{figure}[!h]
 \centering
 \includegraphics[width=5.6cm, height=3.3cm]{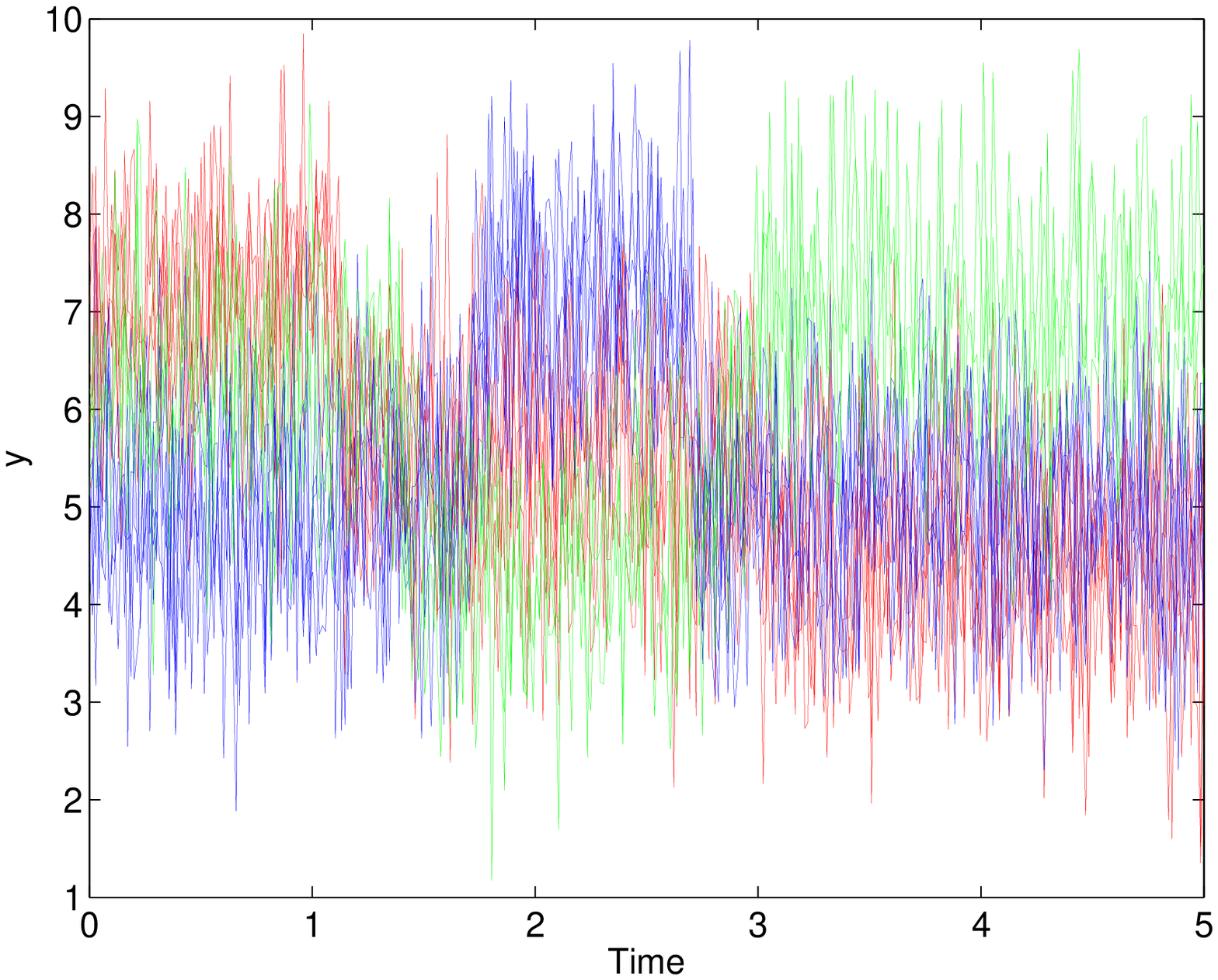} \\
 \includegraphics[width=5.6cm, height=4.3cm]{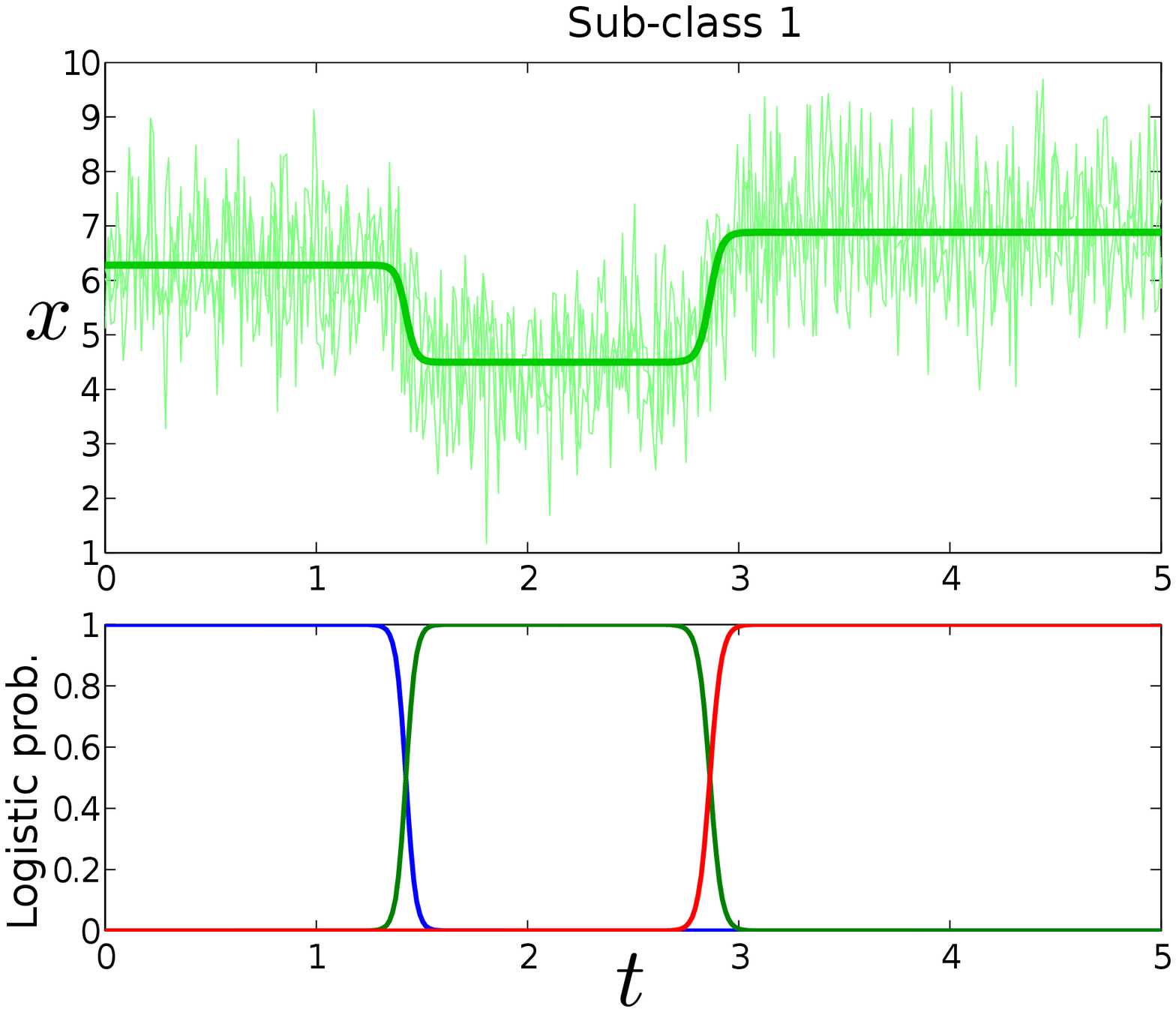} 
 \includegraphics[width=5.6cm, height=4.3cm]{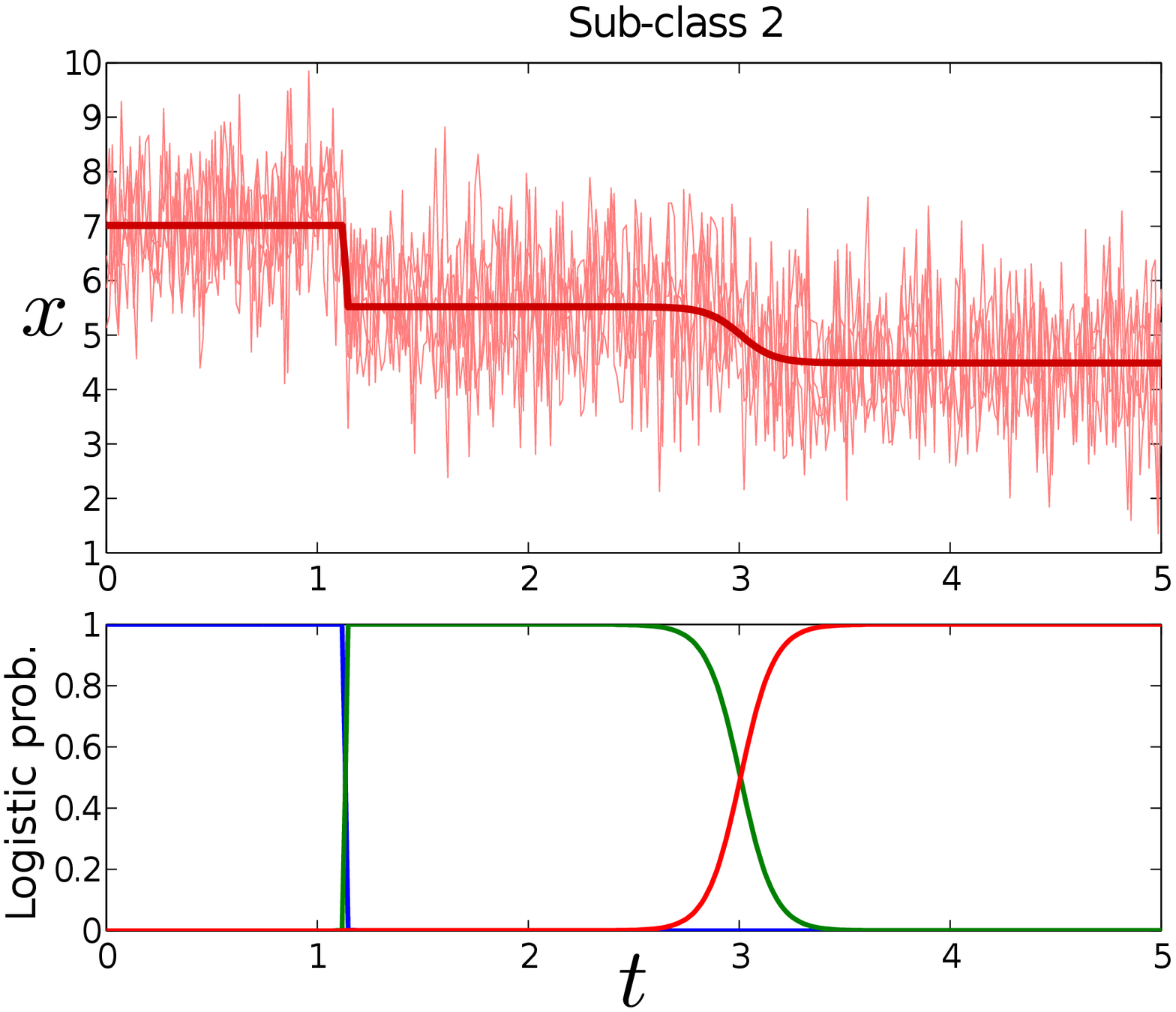}
 \includegraphics[width=5.6cm, height=4.3cm]{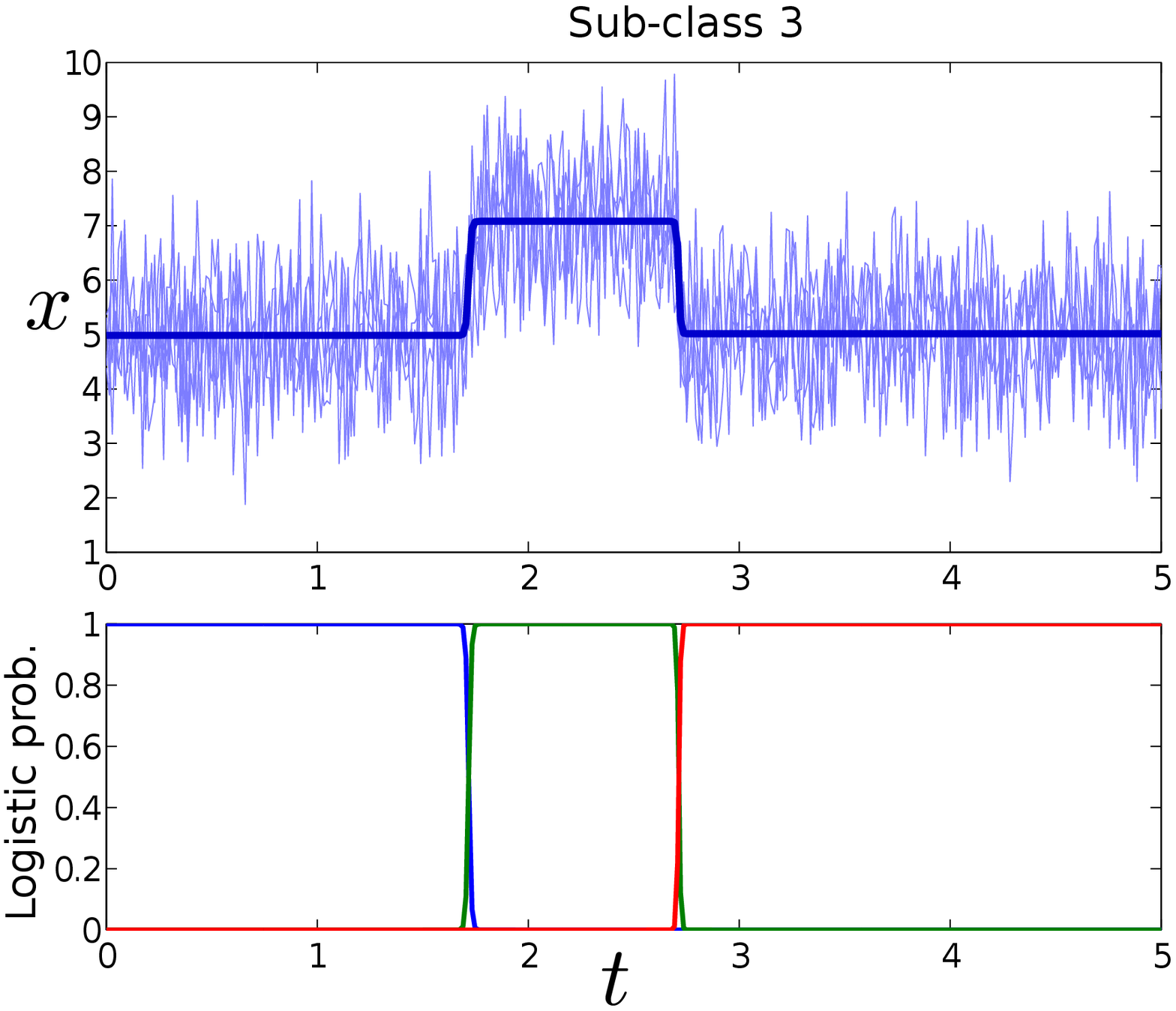}
\caption{\label{fig: results for the complex-shaped class}
The estimated sub-classes colored according to the partition given by the EM algorithm for the proposed approach (top); Then are presented separately each sub-class of curves with the estimated mean curve in bold line (top sub-plot) and the corresponding logistic probabilities that govern the hidden regimes (bottom sub-plot).} 
\end{figure}

Figure \ref{fig: results for the complex-shaped class} shows the obtained modeling results for the complex-shaped class shown in Figure \ref{fig: simulated complex shaped class}. First, it can be observed that the proposed unsupervised approach accurately decomposes the class into homogeneous sub-classes of curves. It can also be observed that the approach is able to automatically determine the underlying hidden regimes for the sub-classes. Furthermore, the flexibility of the logistic process used to model the hidden regimes allows for accurately approximating both abrupt and/or smooth regime changes within each sub-class. This can be clearly seen on the logistic probabilities which vary over time according to both which regime is active or not and how is the transition from one regime to another over time (i.e., abrupt or smooth transition from one regime to another). It can also be noticed that, approximating this class with a single mean curve, which is the case when using FLDA, fails; the class is clearly heterogeneous. Using FMDA based on polynomial or spline regression mixture (i.e., FMDA-PRM or FMDA-SRM) does not provide significant modeling improvements since, as we can clearly see on the data, the subclasses present abrupt and smooth regime changes for which these two approaches are not well adapted. This can be observed on the obtained results of mean intra-class inertia  given in Table \ref{table: results for simulated curves}. 

Table \ref{table: results for simulated curves} also shows the misclassification error rates 
 obtained with the proposed FMDA-MixRHLP approach and alternative approaches.   
\begin{table}[!h]
\centering
\begin{tabular}{|l|c|c|} 
\hline
Approach &  Classif. error rate (\%) & Intra-class inertia\\ 
\hline
\hline
FLDA-PR   & 21   &  $7.1364 \times 10^3$ \\
FLDA-SR  & 19.3 &  $6.9640 \times 10^3$ \\
FLDA-RHLP & 18.5 &  $6.4485\times 10^3$ \\
\hline
FMDA-PRM & 11 &    $6.1735 \times 10^3$\\
FMDA-SRM     & 9.5 & $5.3570 \times 10^3$ \\
{\bf FMDA-MixRHLP } & 5.3 & $3.8095\times 10^3$ \\
\hline
\end{tabular}
\caption{\label{table: results for simulated curves}
Obtained results for the simulated curves.}
\end{table} 
As expected, it can be seen that the FMDA approaches provide better results compared to FLDA approaches. This is due to the fact that using a single model for complex-shaped classes  (i.e., when using FLDA approaches) is not adapted. It can also be observed that the proposed functional mixture discriminant approach based on hidden logistic process regression (FMDA-MixRHLP) outperforms the alternative FMDA based on polynomial regression mixtures (FMDA-PRM) or spline regression mixtures (FMDA-SRM). This performance is attributed to the flexibility of the MixRHLP model thanks to the logistic process which is well adapted for modeling the regime changes. 

In the second situation, the proposed approach  is applied 
on the waveform curves of Breiman \cite{breiman}.
\subsubsection{Waveform curves of Breiman}
\label{sssec: experiments using waveform data}

\begin{figure}[!h]
 \centering
\includegraphics[width=4.27cm,height=3.3cm]{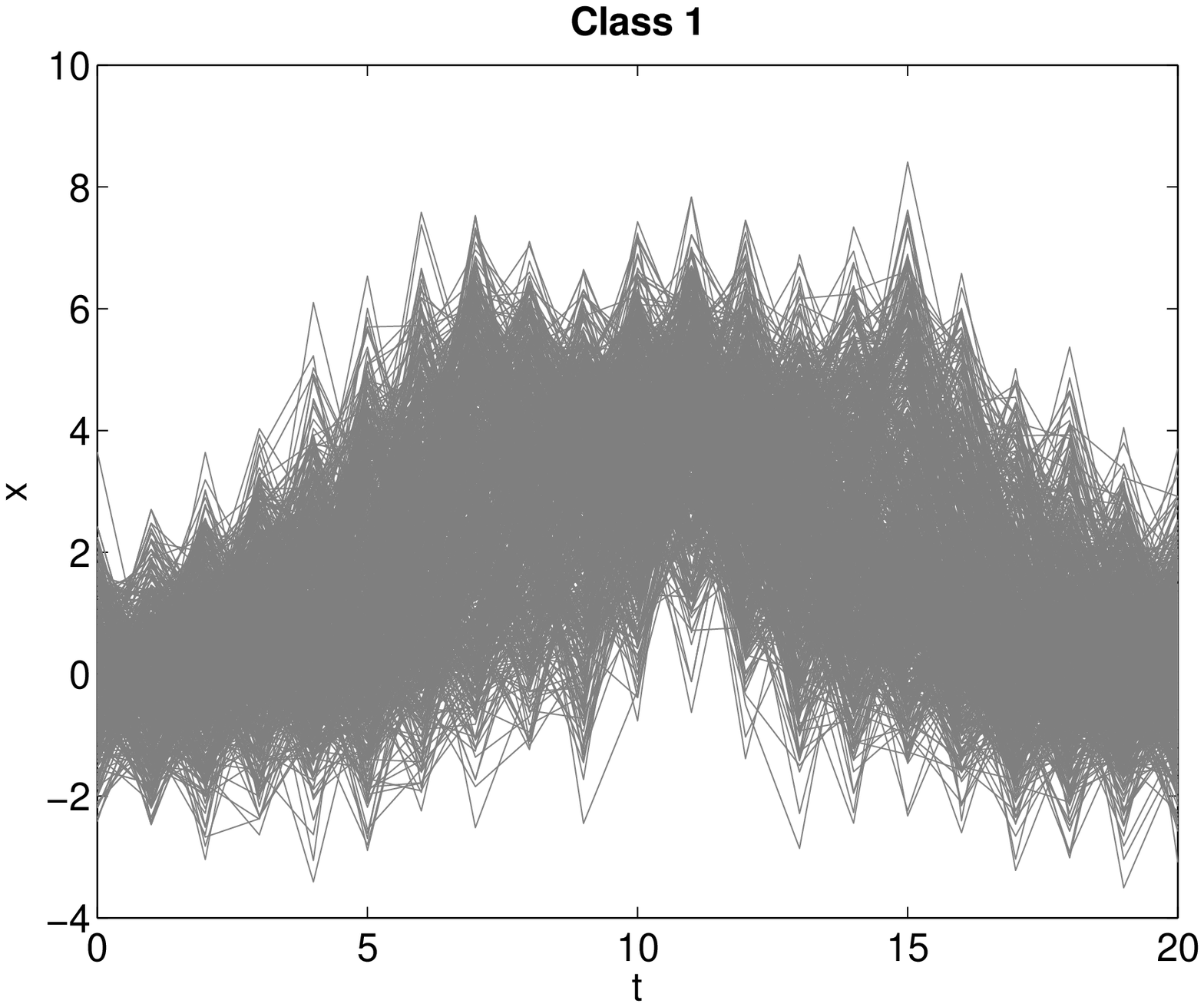}
\includegraphics[width=4.27cm,height=3.3cm]{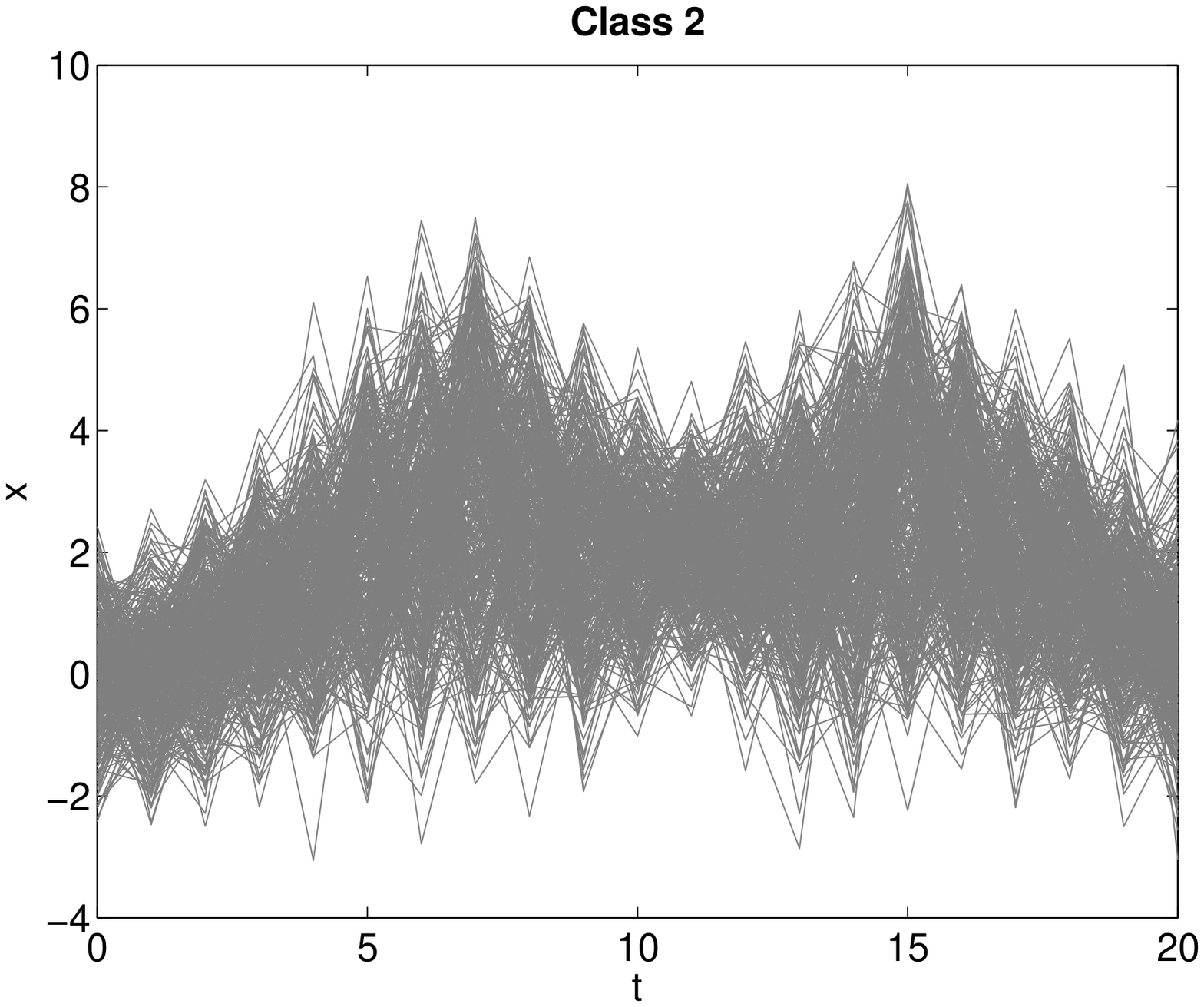}
\includegraphics[width=4.27cm,height=3.3cm]{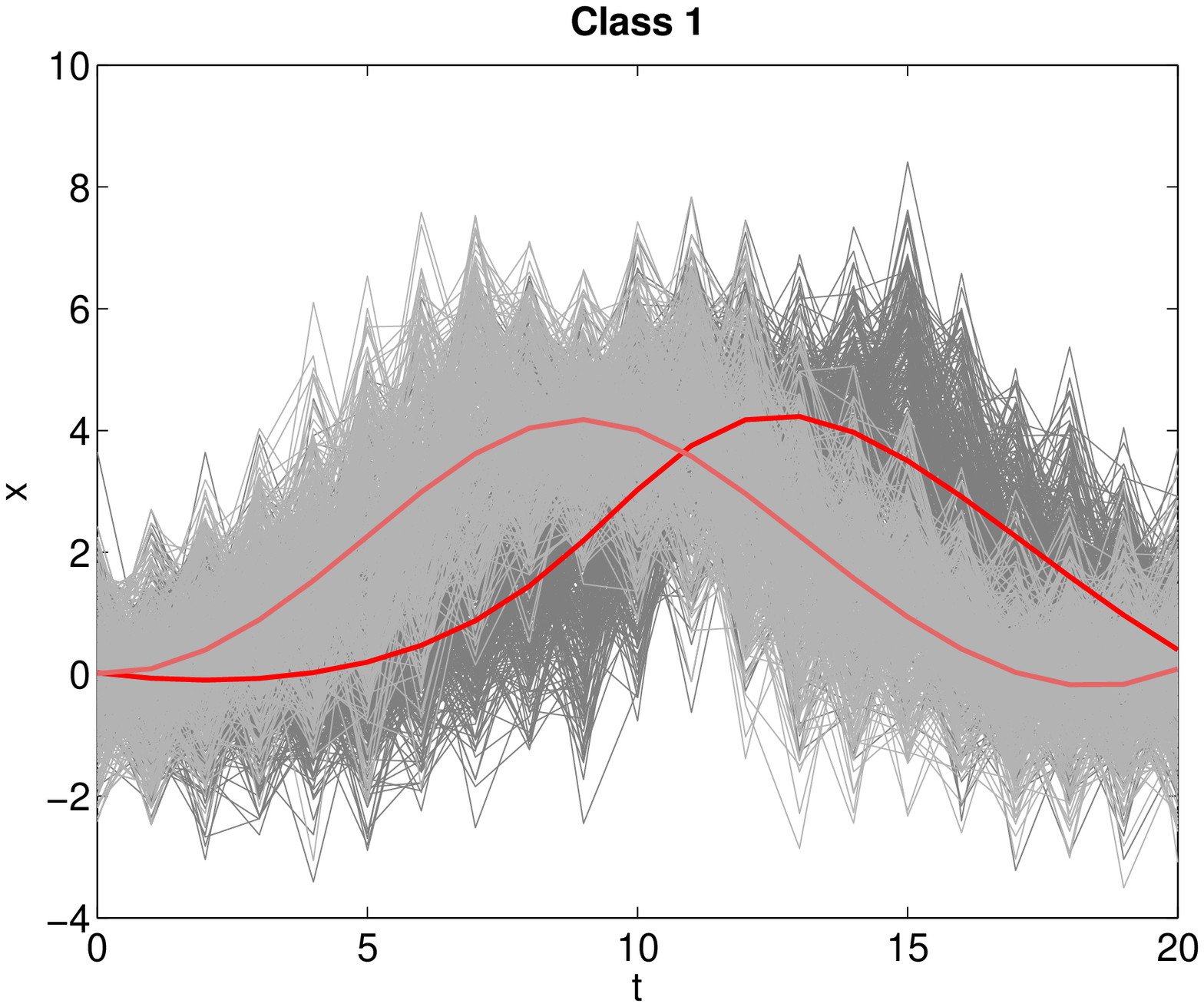}
\includegraphics[width=4.27cm,height=3.3cm]{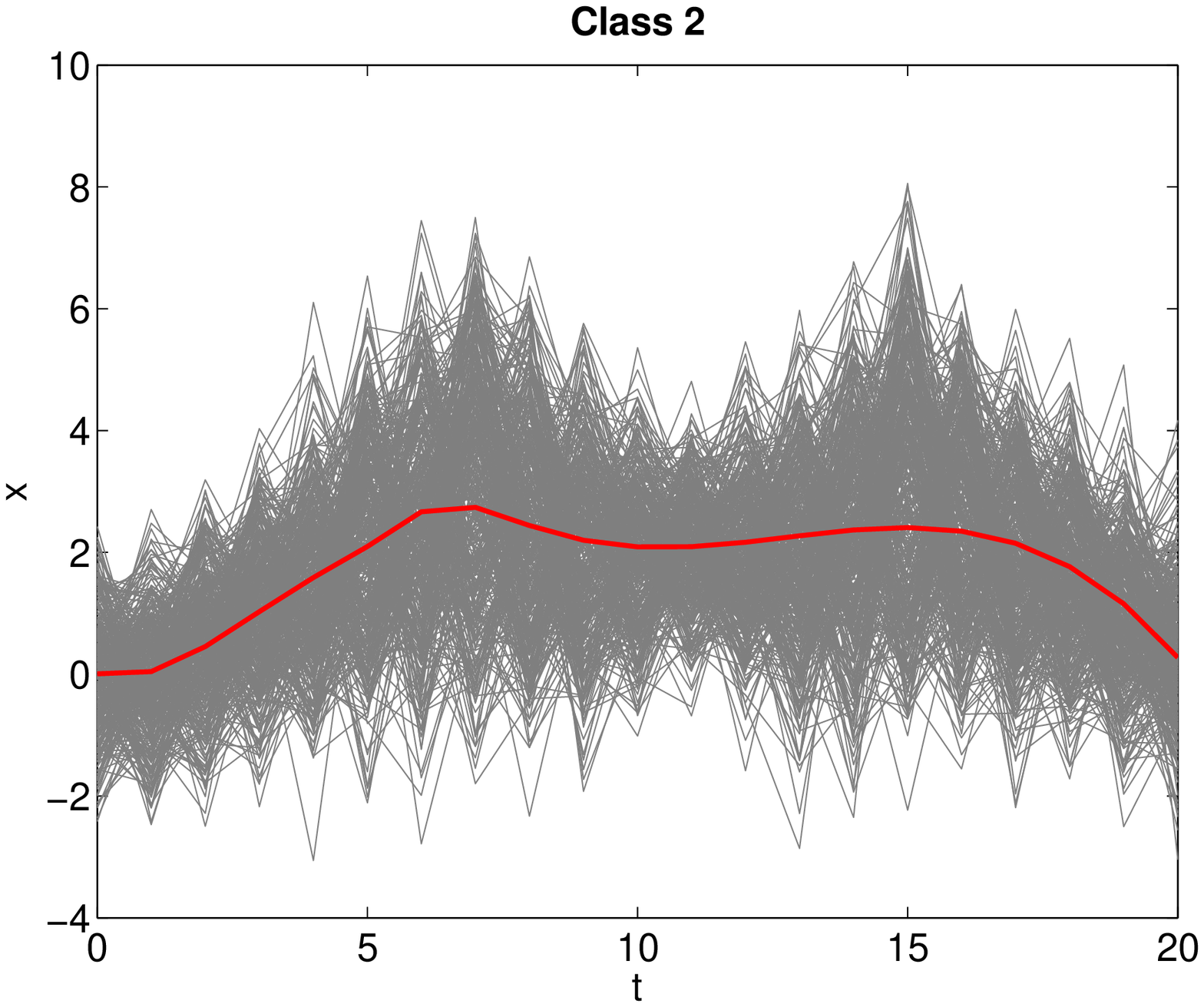}
\includegraphics[width=4.27cm,height=3.3cm]{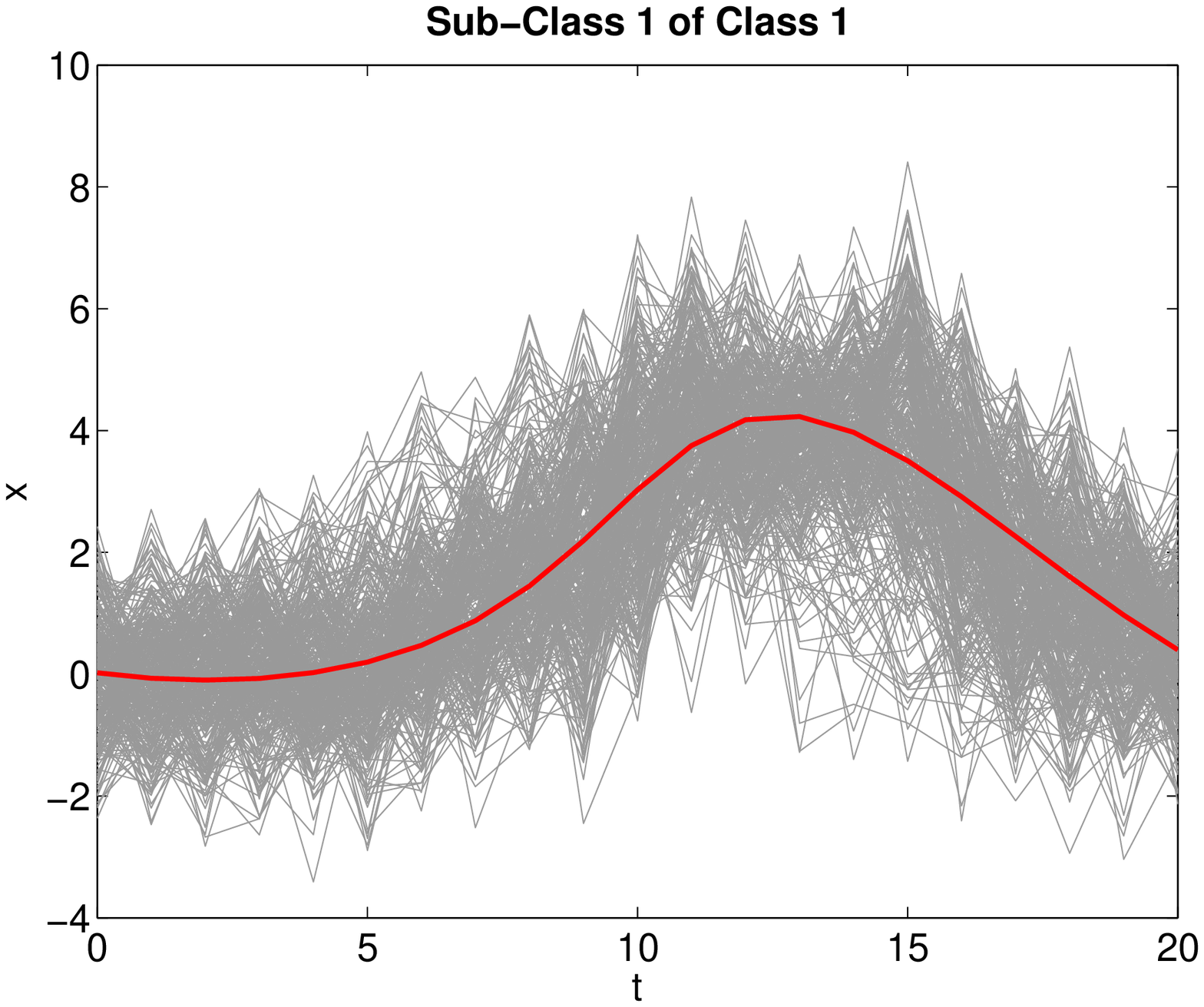}
\includegraphics[width=4.27cm,height=3.3cm]{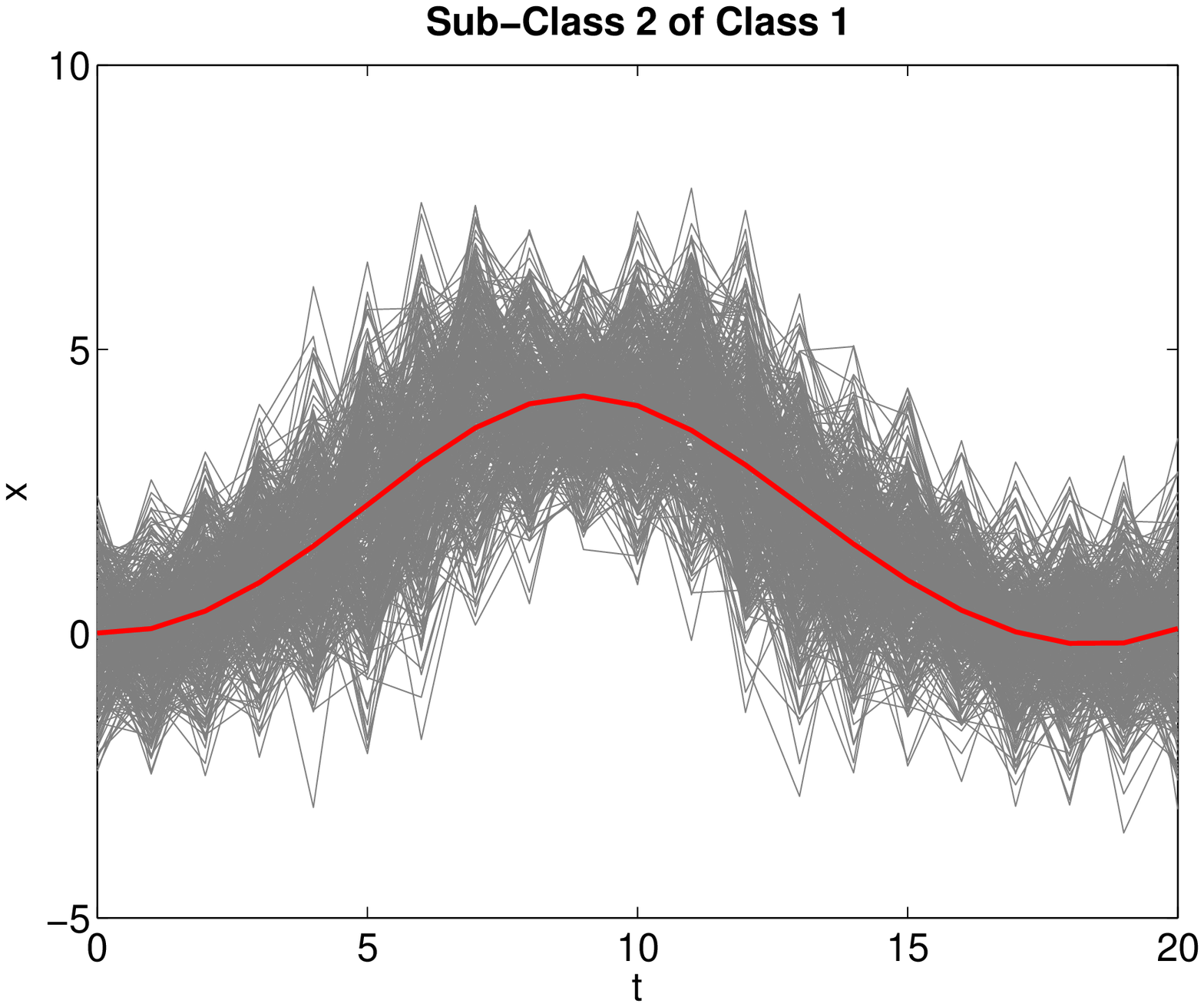} 
  \caption{Modeling results for the waveform curves: (top) the waveforms (500 curves per class) where the first class is composed of two sub-classes, (middle) the waveforms and the estimated subclasses for class 1 and the corresponding mean curves for each class, and (bottom) the two subclasses of class 1 shown separately with their corresponding mean curves.}
 \label{fig: waveform curves examples and estimations}
\end{figure}

The waveform data introduced by \cite{breiman} 
consist of a three-class problem where each curve is generated as follows:
\begin{itemize}
\item $\bx_i(t)=uf_1(t) + (1-u)f_2(t) + \epsilon_t$ for the class 1;
\item $\bx_i(t)=uf_2(t) + (1-u)f_3(t) + \epsilon_t$ for the class 2;
\item $\bx_i(t)=uf_1(t) + (1-u)f_3(t) + \epsilon_t$ for the class 3.
\end{itemize}
where $u$ is a uniform random variable on $(0,1)$, \linebreak
$f_1(t)=\max (6-|t-11|,0)$;
$f_2(t)=f_1(t-4)$;
 $f_3(t)=f_1(t+4)$
and $\epsilon_t$ is a zero-mean Gaussian noise with unit standard deviation.
The temporal interval considered for each curve is $[0;20]$ with a constant period of sampling of 1 second. 
For the experiments considered here, inorder to have a heterogeneous class, 
we combine both class 1 and class 2 to form a single class called class 1. Class 2 will therefore used to refer to class 3 in the previous description of the waveform data. Figure \ref{fig: waveform curves examples and estimations} (top) shows curves from the two classes.

Figure \ref{fig: waveform curves examples and estimations} (middle) shows the obtained modeling results for each of the two classes by applying the proposed approach. We can see that the two sub-classes for the first classes are well identified. These two sub-classes (clusters) are shown separately on Figure \ref{fig: waveform curves examples and estimations} (bottom) with their corresponding mean curves. We notice that for this data set, all FMDA approaches provide  very similar results regarding both the classification and the approximation since, as it can be seen, the complexity for this example is only related to the dispersion of the first class into sub-classes, and there are no explicit regime changes; each sub-class can therefore also be accurately approximated by a polynomial or a spline function.  %

\subsubsection{Experiments on real data}

\begin{figure}[!h] 
\includegraphics[width=4.2cm,height=3.2 cm]{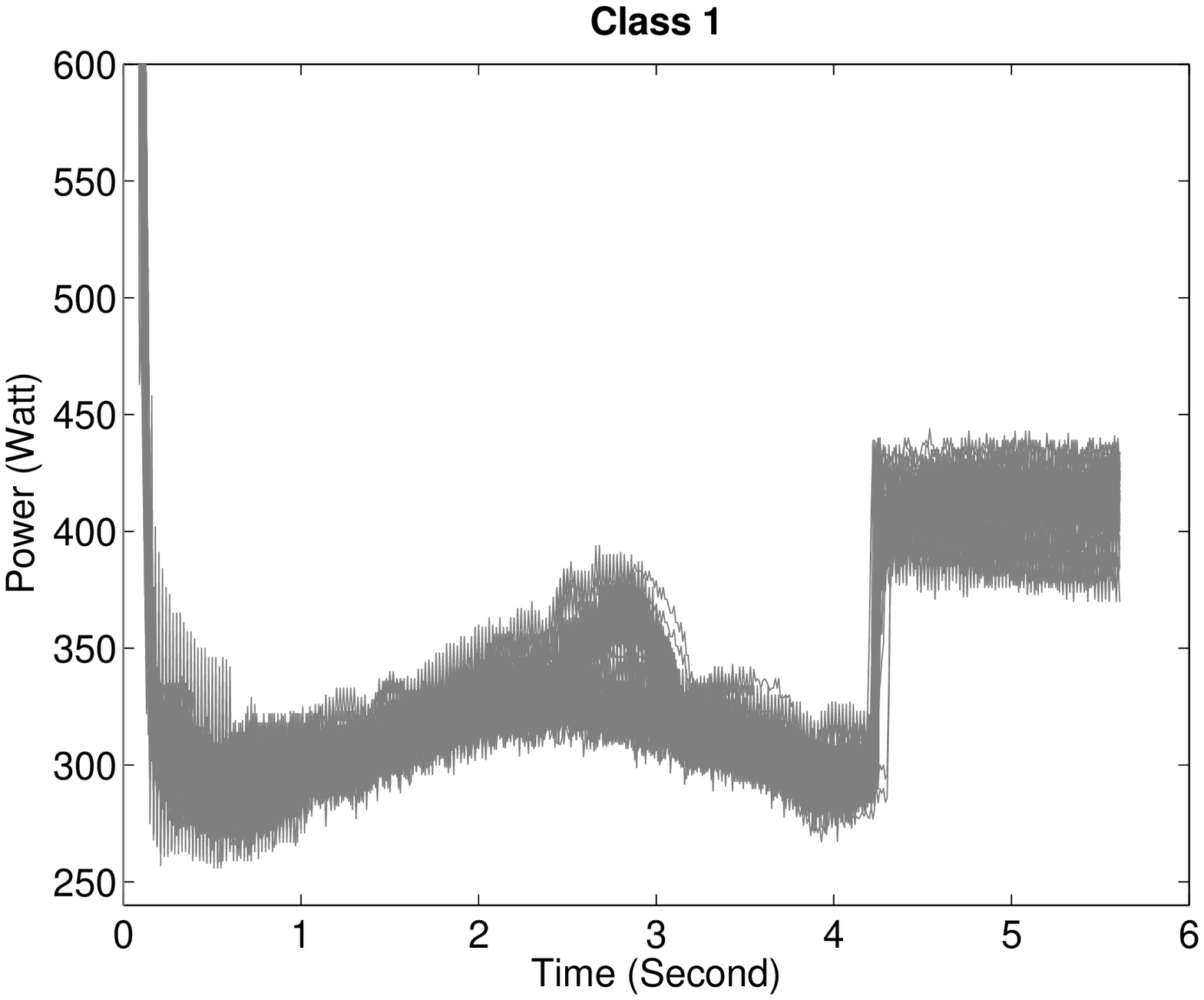}
\includegraphics[width=4.2cm,height=3.2 cm]{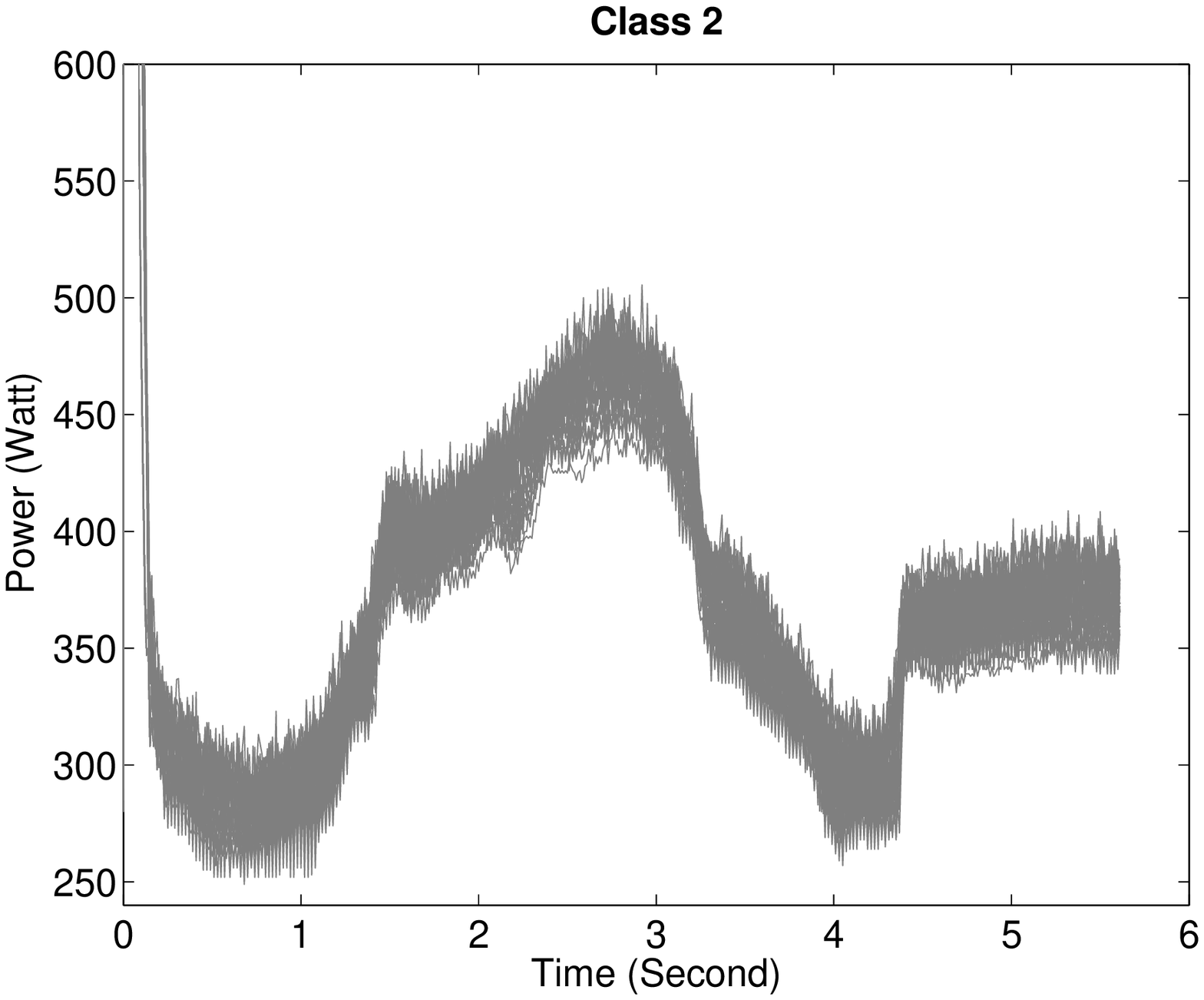} 
\caption{75 switch operation curves from the first class (left) and 45 curves from the second class (right).}
\label{fig: switch-curves-classes}
\end{figure}

\begin{figure}[!h]
\centering
\includegraphics[width=4.27cm,height=3 cm]{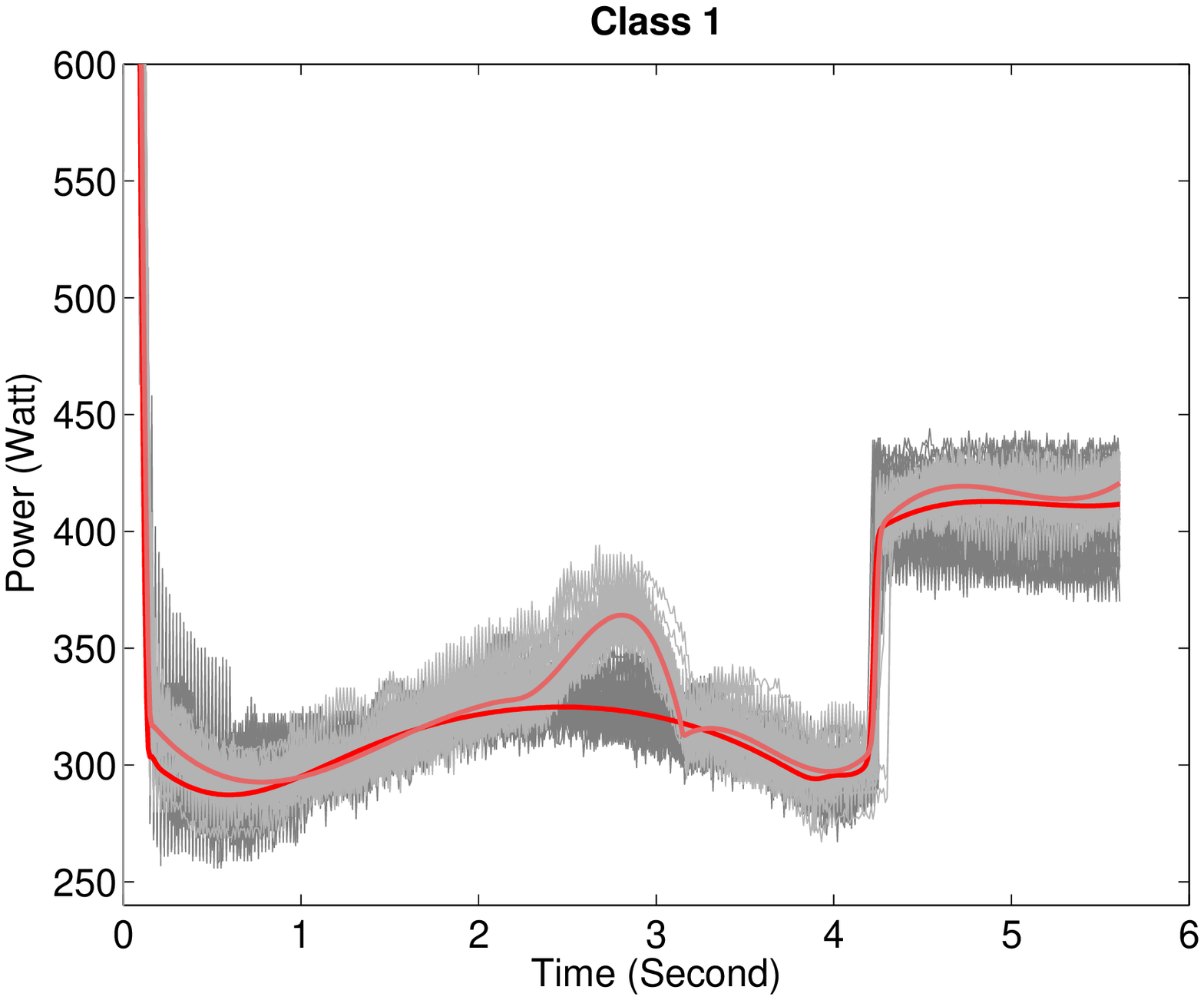}
\includegraphics[width=4.27cm,height=3 cm]{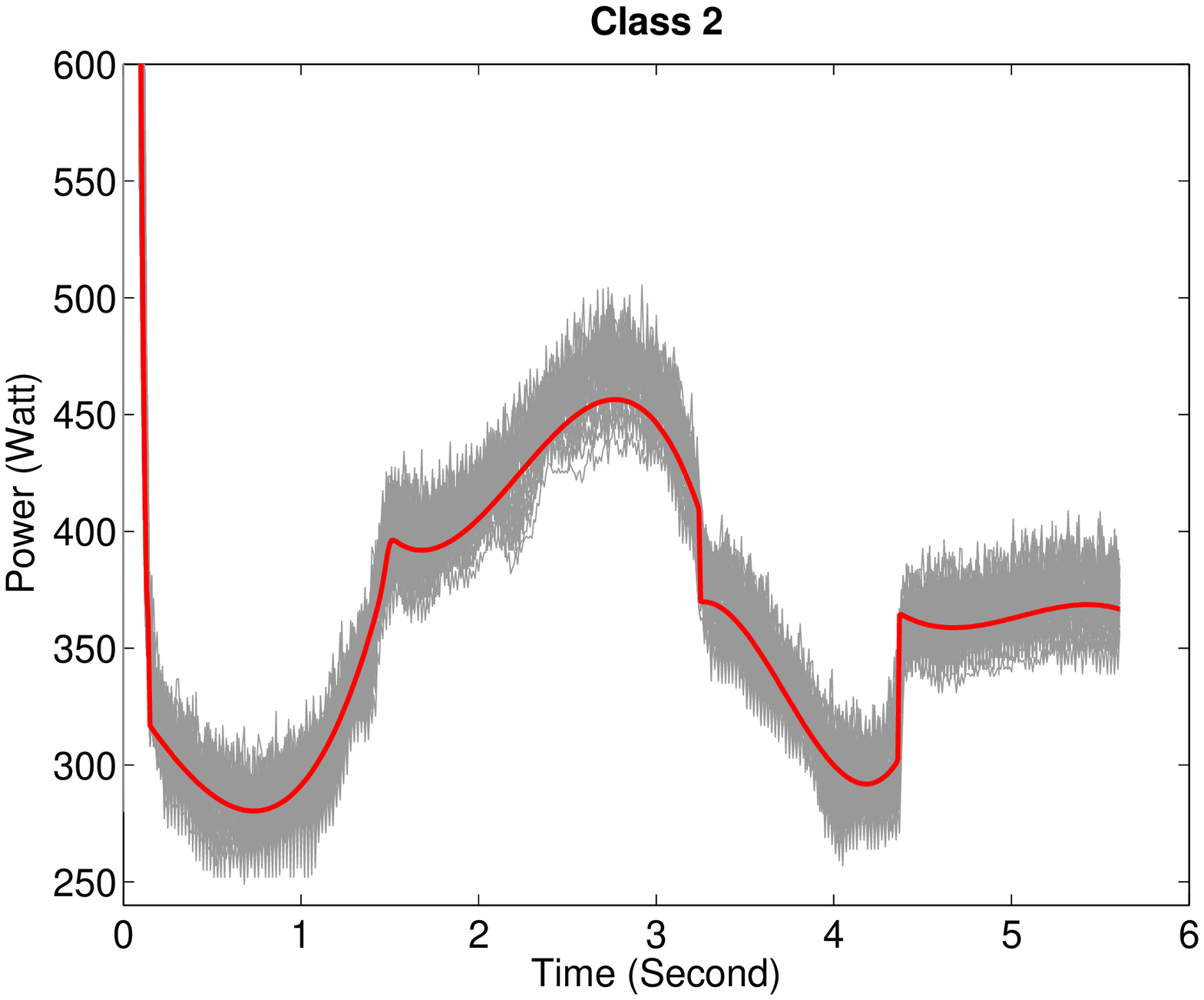}
\includegraphics[width=4.27cm,height=3 cm]{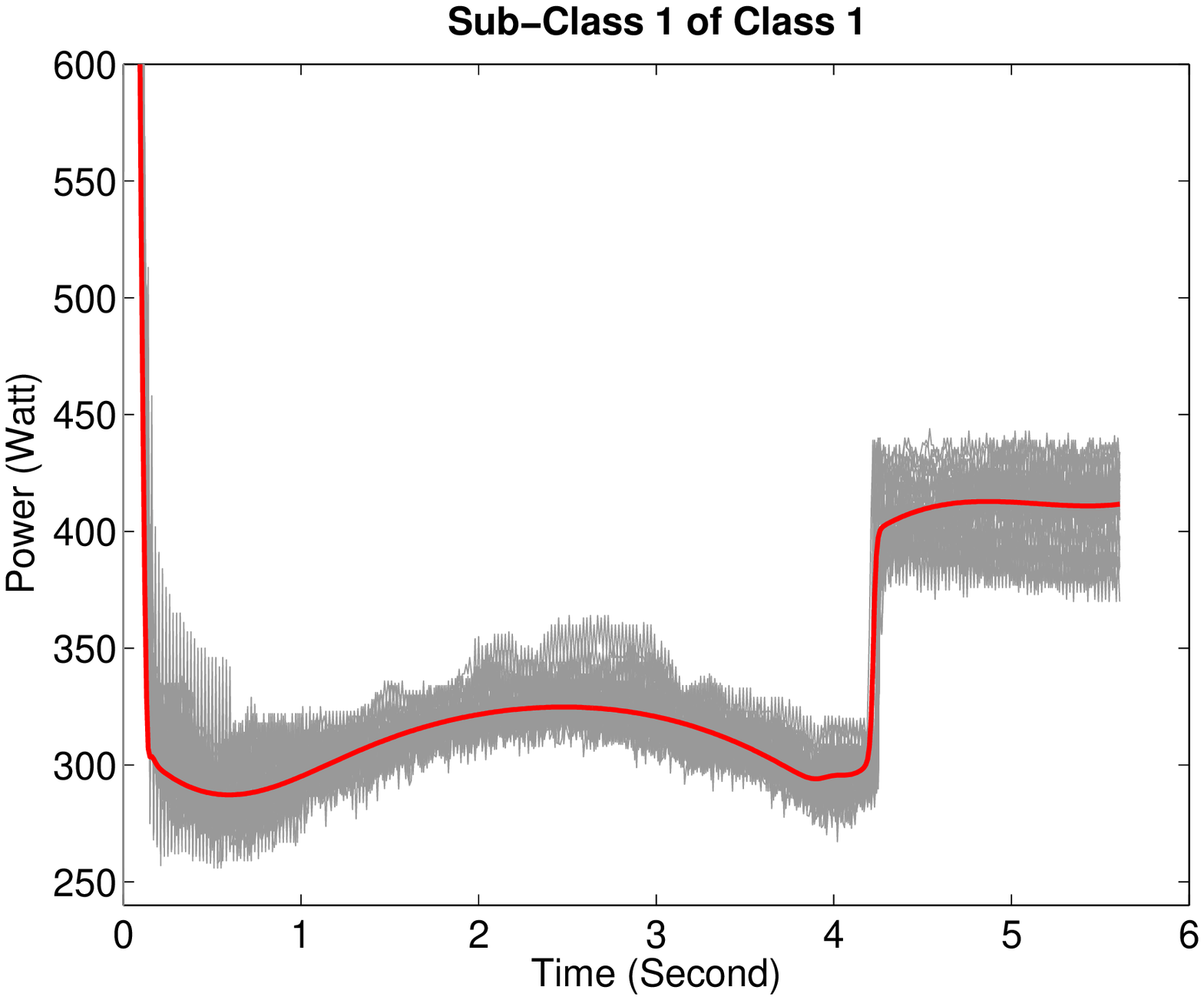}
\includegraphics[width=4.27cm,height=3 cm]{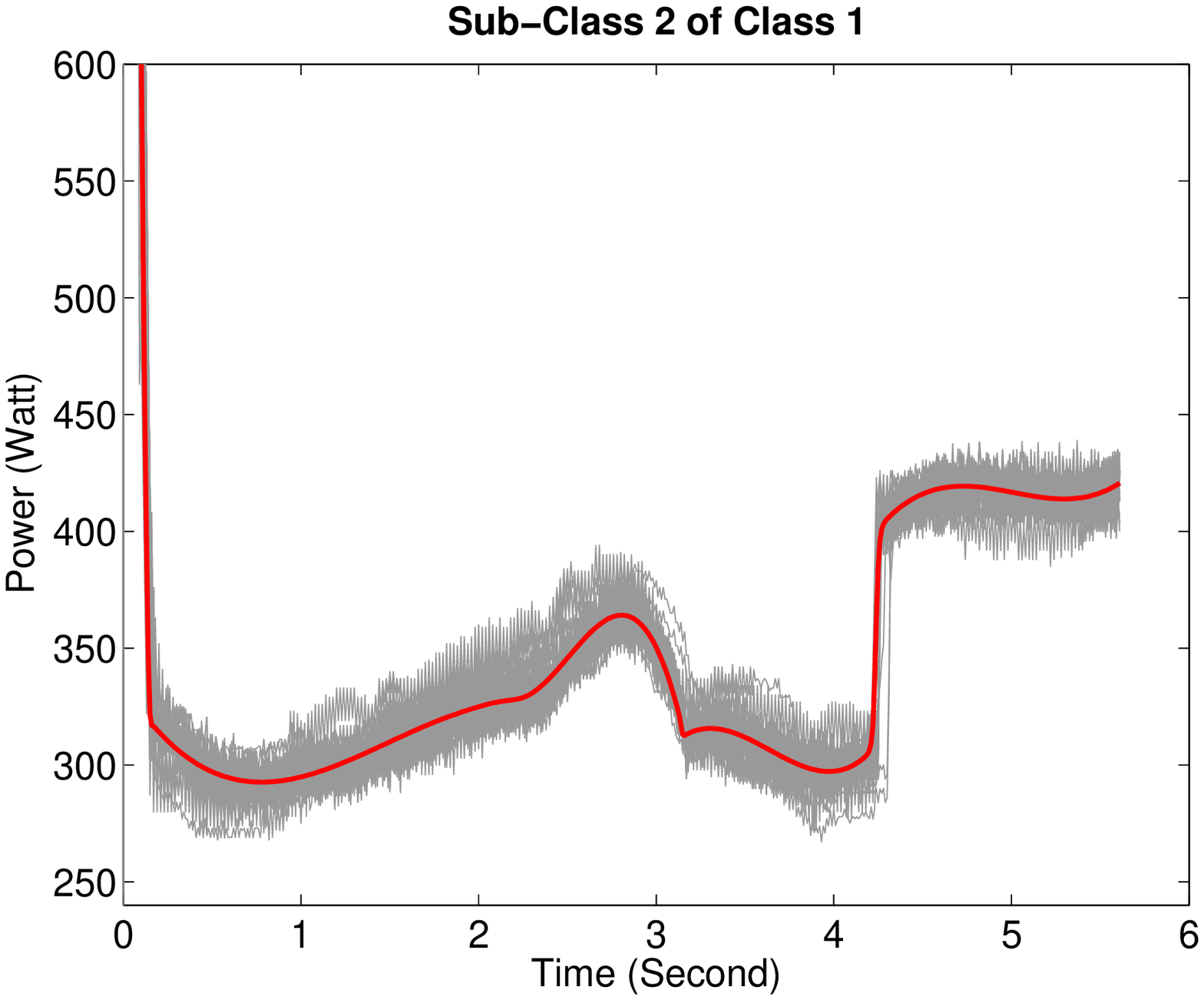}
\includegraphics[width=4.27cm,height=3 cm]{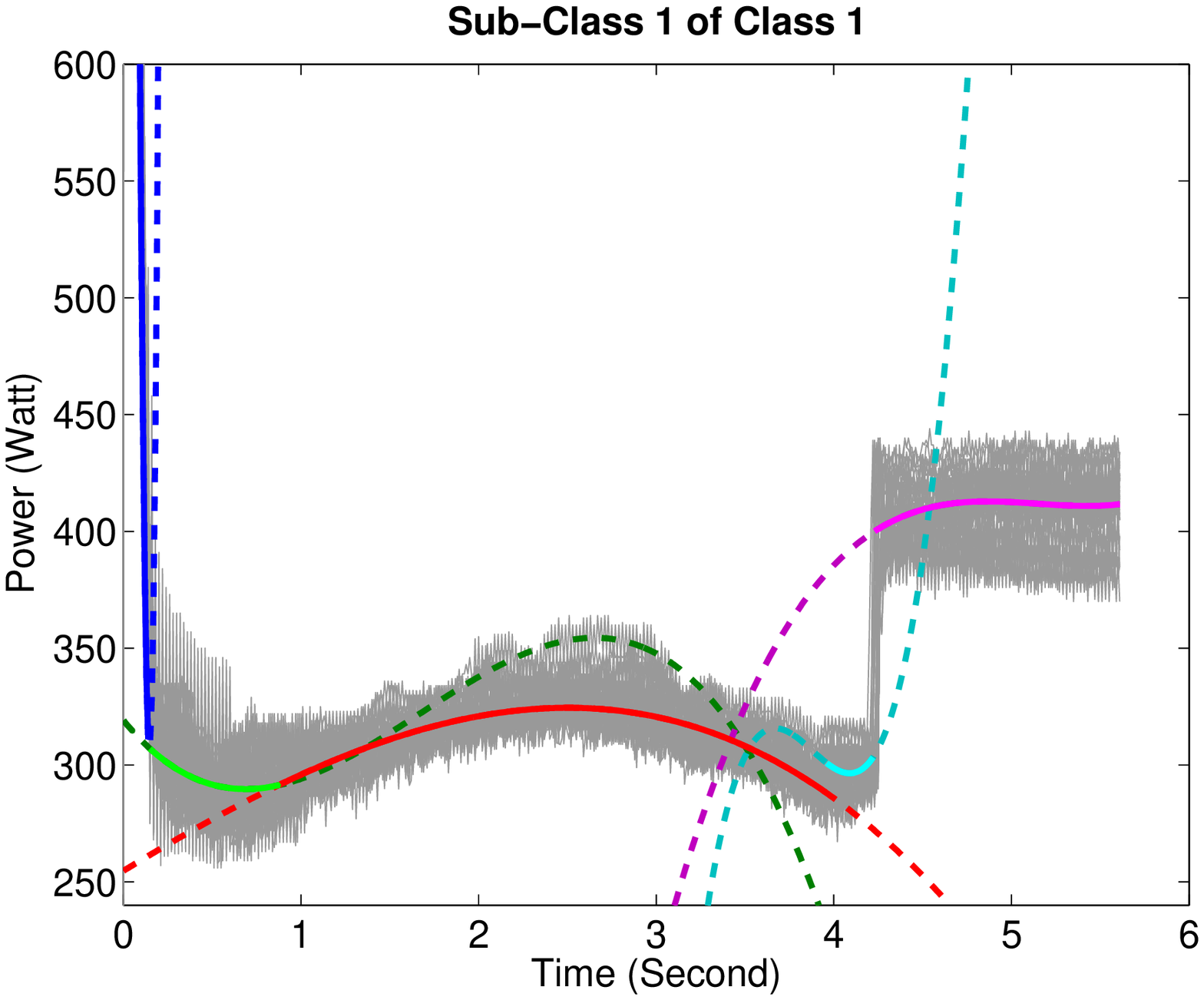}
\includegraphics[width=4.27cm,height=3 cm]{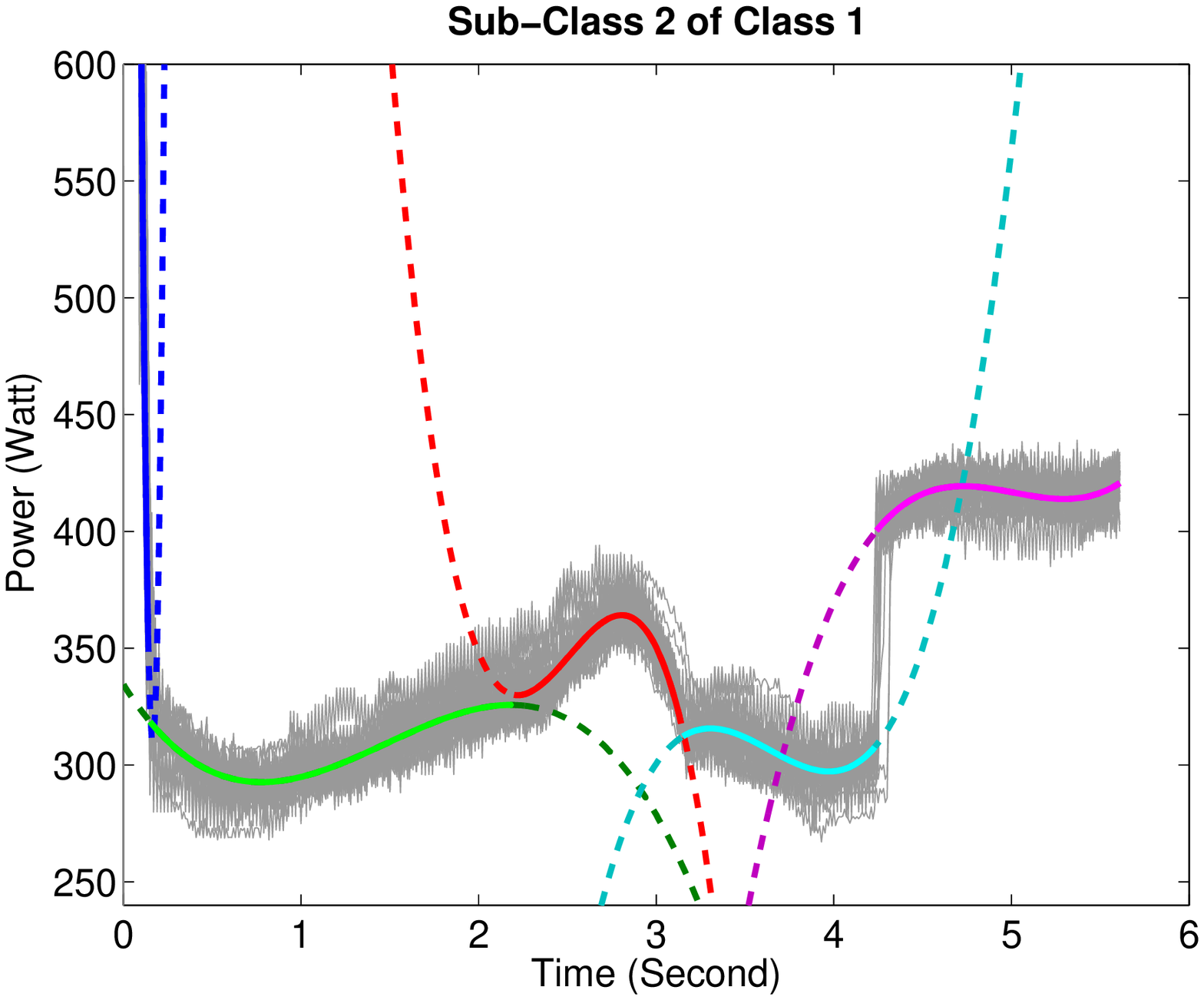}
\includegraphics[width=4.27cm,height=2.8 cm]{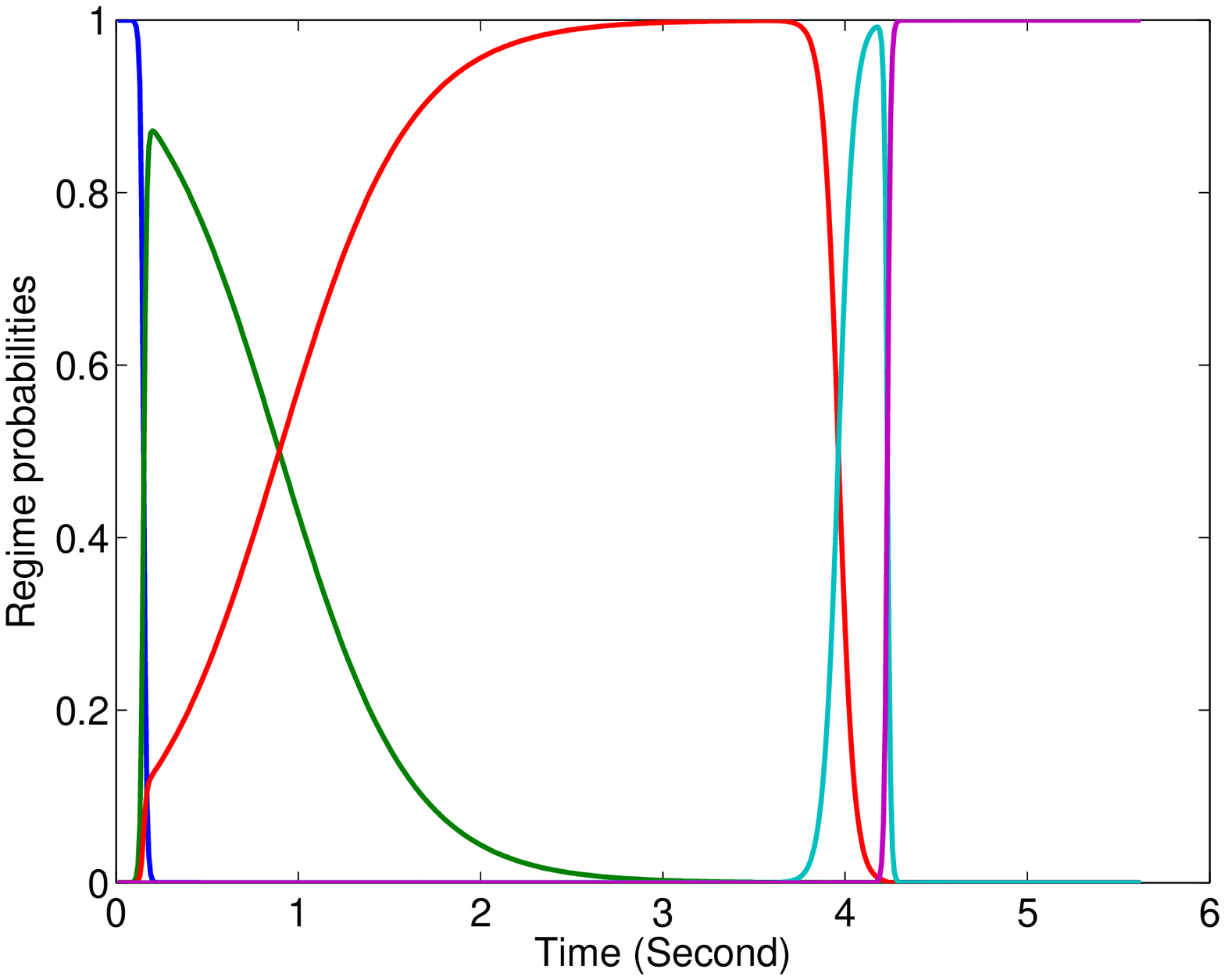}
\includegraphics[width=4.27cm,height=2.8 cm]{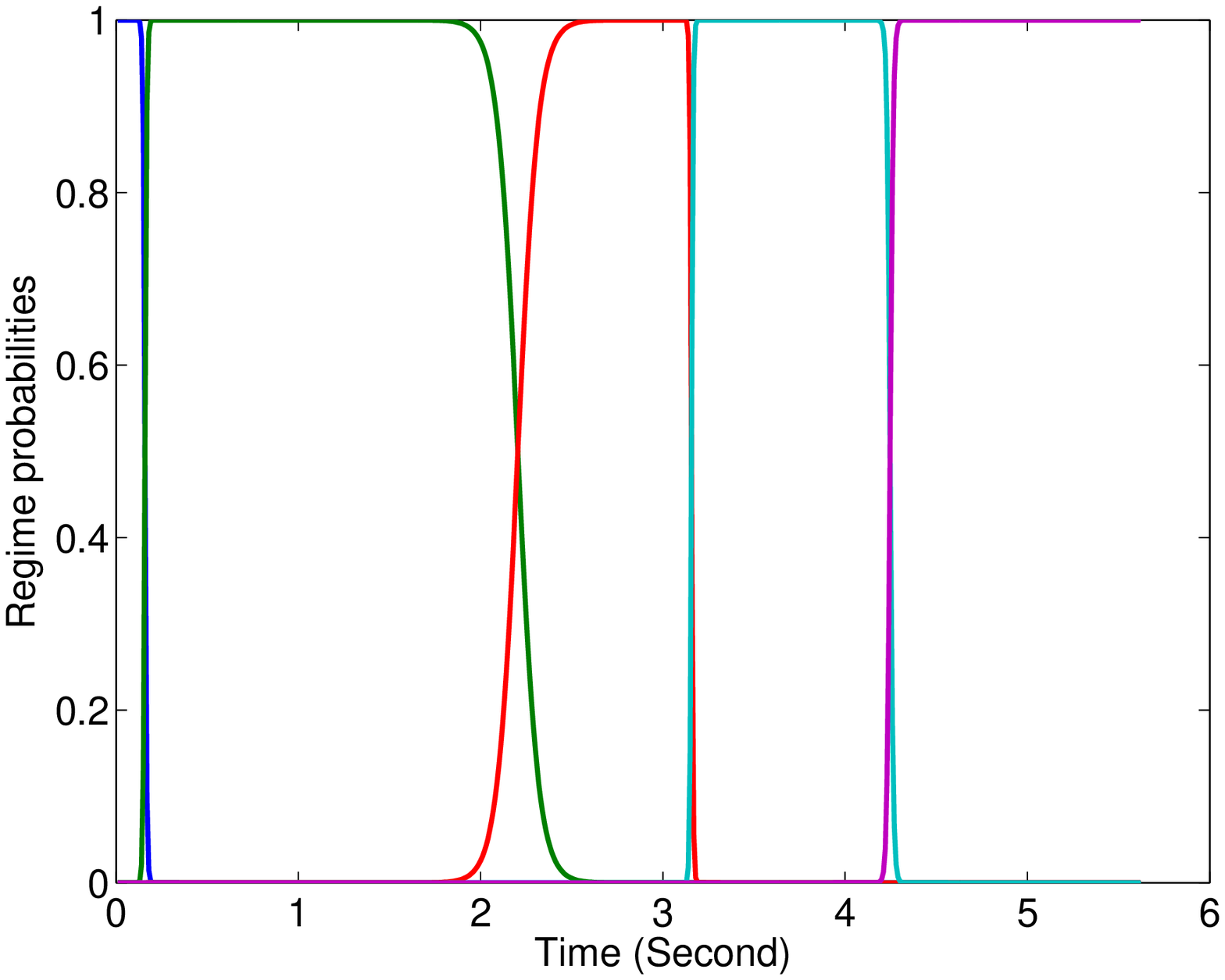}
\includegraphics[width=4.27cm,height=3 cm]{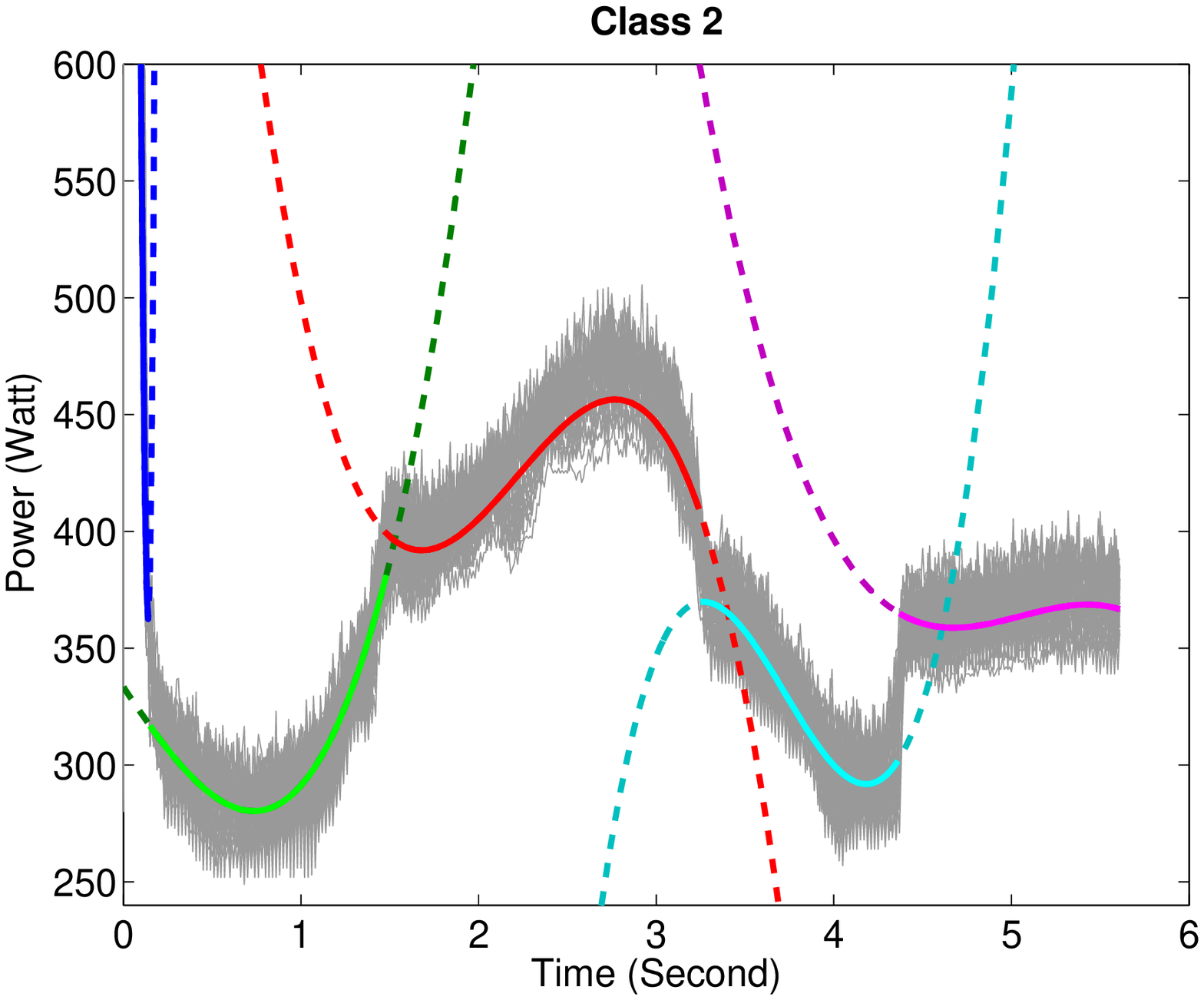}
\includegraphics[width=4.27cm,height=2.9 cm]{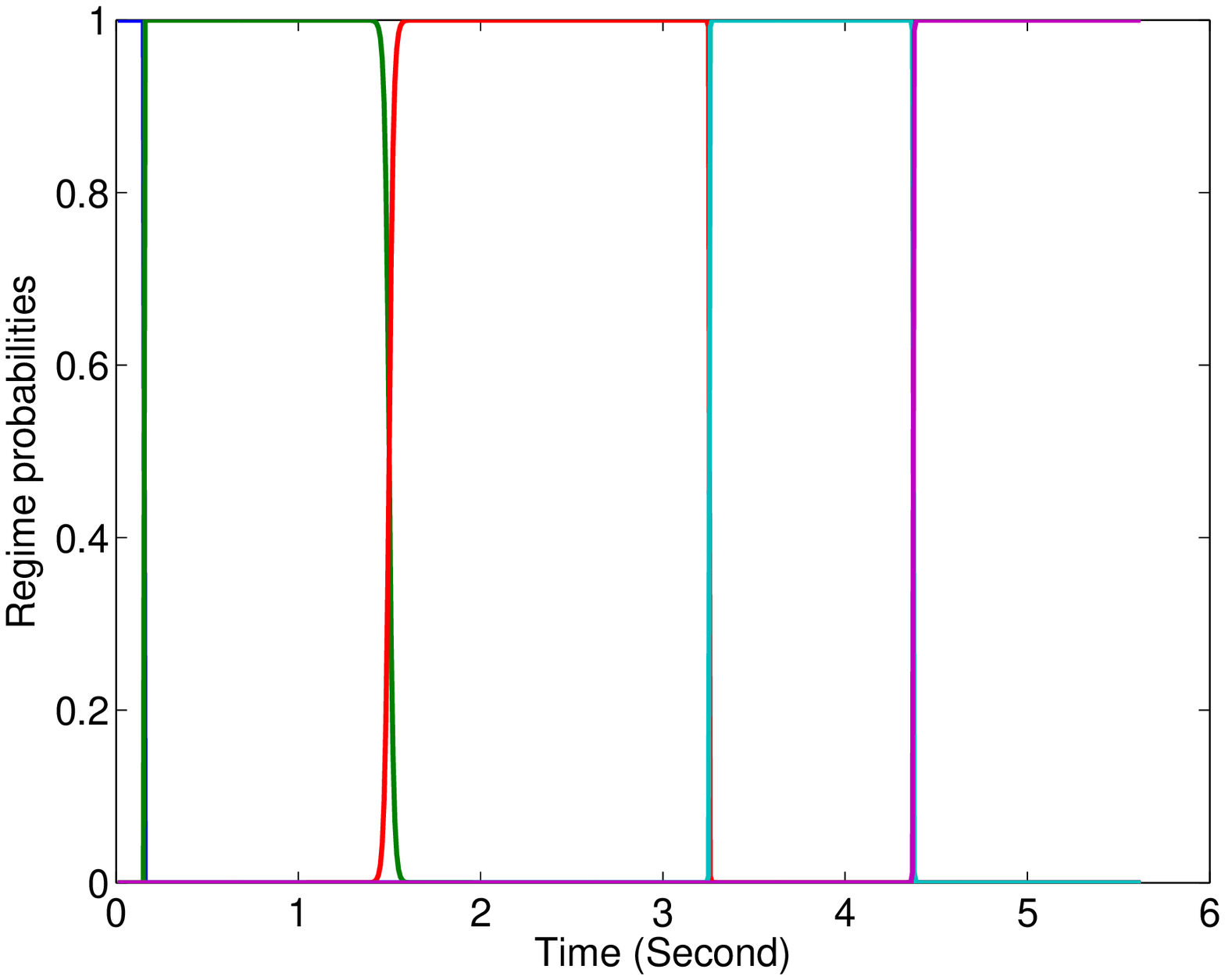}\\  
\caption{\label{fig: switch-curves-MixRHLP results}
Results obtained with the proposed model  for the real curves.
The estimated sub-classes for class 1 (top-left)  and the corresponding mean curves (top) provided by the proposed approach; Then, we show  separately each sub-class of class 1 with the estimated mean curve presented in a bold line (top sub-plot), the polynomial regressors (degree $p=3$), the corresponding logistic proportions that govern the hidden process, and finally in the bottom plots we show the same results for class 2.} 
\end{figure}

 In this section, we use a database issued from a railway diagnosis application as studied in \cite{chamroukhi_et_al_neurocomputing2010}\cite{chamroukhi_et_al_NN2009}\cite{chamroukhi_adac_2011}. This database is composed of $120$ labeled real switch operation curves. In \cite{chamroukhi_et_al_neurocomputing2010}\cite{chamroukhi_et_al_NN2009}\cite{chamroukhi_adac_2011}, the data were used to perform classification into three classes : no defect, with a minor defect and with a critical defect. In this study, we rather consider two classes where the first one is composed by the curves with no defect and with a minor defect so that the decision will be either with or without defect.  The goal is therefore to provide an accurate automatic modeling especially for Class 1 which is henceforth dispersed into two sub-classes. The cardinal numbers of the classes are $n_1=75$ and  $n_2=45$ respectively. Figure \ref{fig: switch-curves-classes} shows each class of curves, where the first class is composed of two sub-classes.
Figure \ref{fig: switch-curves-MixRHLP results} shows the modeling results provided by the proposed approach for each of the two classes. It shows the two sub-classes estimated for class 1 and the corresponding mean curves for the two classes. We also present the estimated polynomial regressors for each set of curves and the corresponding probabilities  of the logistic process that govern the regime changes over time. We see that the proposed method ensure both a decomposition of the complex shaped class into sub-classes and at the same time, a good approximation of the underlying regimes within each homogeneous set of curves. Indeed, it can be seen that the logistic process probabilities are close to $1$ when the $r$th regression model seems to be the best fit for the curves and vary over time according to the smoothness degree of regime transition. 
Then, the obtained classification results, by considering the FLDA approaches and the FMDA approaches (which are more competitive) and gave the best results for simulations, are given in Table \ref{table: results for switch curves}. 
{\small \begin{table}[!h]
\centering
\begin{tabular}{|l|c|c|} 
\hline
Approach &  Classif. error rate (\%) & Intra-class inertia\\ 
\hline 
\hline 
FLDA-PR   & 11.5   &  $10.7350 \times 10^9$ \\
FLDA-SR  & 9.53 &  $9.4503 \times 10^9$ \\
FLDA-RHLP & 8.62 &  $8.7633\times 10^9$ \\
\hline 
FMDA-PRM 	& 9.02 &    $7.9450 \times 10^9$\\
FMDA-SRM     & 8.50 & $5.8312 \times 10^9$ \\
{\bf FMDA-MixRHLP } & 6.25 & $3.2012 \times 10^9$ \\
\hline
\end{tabular}
\caption{\label{table: results for switch curves}
Obtained results for the real curves.}
\end{table}}

We can see that, although the classification results are similar for the FMDA approaches, the difference in terms of curves modeling (approximation) is significant, for which the proposed approach clearly outperforms the alternatives. This is attributed to the fact that the use of polynomial regression (mixtures) or spline regression (mixtures) does not fit at best the regime changes compared to the proposed model. Finally we notice that the proposed algorithm converges in approximatively  80 iterations.
 
\section{Conclusion}
\label{sec: conclusion}
In this paper, we presented a new model-based approach for functional data classification. It uses a specific functional mixture discriminant analysis incorporating a hidden process regression model, particularly adapted for modeling complex-shaped classes of curves presenting regime changes. 
The parameters of each class are estimated  in an unsupervised way by a dedicated EM algorithm.  The experimental results on simulated data and real data demonstrated the benefit of the proposed approach  
 as compared to existing alternative functional discriminant methods.  
Future work will concern experiments on additional real data including time course gene expression curves; 
 We also plan to investigate more model selection approaches which have been shown to perform better then BIC in the case of finite mixture models, such as the one proposed in \cite{BouguilaZ07}. 
 We will as well investigate Bayesian learning techniques from functional data to explicitly incorporate some prior knowledge on the data structure to better control the model complexity.
%
\bibliographystyle{IEEEtran}
\bibliography{references} 
 
\end{document}